# Progressive Damage Modelling and Fatigue Life Prediction of Plain-weave Composite Laminates with Low-velocity Impact Damage

**Zheng-Qiang Cheng[a,b], Wei Tan[b], Jun-Jiang Xiong[a,*]**

[a] School of Transportation Science and Engineering, Beihang University, Beijing 100191, People's Republic of China (* Corresponding author. E-mail address: jjxiong@buaa.edu.cn)

[b] School of Engineering and Materials Science, Queen Mary University of London, London E1 4NS, United Kingdom

**Abstract:** This paper developed a fatigue-driven residual strength model considering the effects of low-velocity impact (LVI) damage and stress ratio. New fatigue failure criteria based on fatigue-driven residual strength concept and fatigue progressive damage model were developed to simulate fatigue damage growth and predict fatigue life for plain-weave composite laminates with LVI damage. To validate the proposed model, LVI tests of plain-weave glass fibre reinforced polymer 3238A/EW250F laminates were conducted, followed by post-impact constant amplitude tension-tension, compression-compression fatigue tests and multi-step fatigue tests. Experimental results indicate that the LVI damage degrades fatigue strength of plain-weave glass fibre composite laminate drastically. The load history also plays an important role on the fatigue accumulation damage of post-impact laminates. The new fatigue progressive damage model achieves a good agreement with fatigue life of post-impact laminates and is able to capture the load sequence effect, opening a new avenue to predict fatigue failure of composite laminates.

## Nomenclature

| | | | |
|---|---|---|---|
| $C, p, q$ | material constants in fatigue residual strength model | $S_{\min,r}$ | minimum value of nominal cyclic stress at arbitrary stress ratio |
| $D$ | Palmgren-Miner cumulative damage | $S_{r_0}$ | maximum absolute value of fatigue stress at specific stress ratio |
| $E$ | Young's modulus for woven ply | $S_{-1}$ | fatigue strength depicted with stress amplitude under fully reversed cyclic loading |
| $f$ | intermediate variable | $S_0$ | fatigue endurance limit |
| $G$ | shear modulus for woven ply | $V_f$ | fibre volume fraction |
| $N$ | fatigue life | $X$ | static strength for woven ply |
| $n$ | number of cyclic loading cycles | $X_0$ | static strength for laminate |
| $r$ | arbitrary stress ratio | $\alpha_0, \alpha_1, \alpha_2$ | model parameters in post-impact static strength model |
| $r_0$ | specific stress ratio | $\beta_0, \beta_1, \beta_2$ | model parameters in post-impact fatigue endurance limit model |
| $R$ | fatigue residual strength | $\nu$ | Poisson's ratio of woven ply |
| $R(n)$ | fatigue residual strength after $n$ number of cycles | $\rho$ | density |
| $S$ | maximum absolute value of fatigue stress | $\sigma$ | stress |
| $S_a$ | amplitude of nominal cyclic stress | $\sigma_{CAI}$ | compression strength after impact |
| $S_m$ | mean of nominal cyclic stress | $\sigma_{TAI}$ | tension strength after impact |
| $S_{max,r}$ | maximum value of nominal cyclic stress at arbitrary stress ratio | $\Pi$ | impact energy per unit thickness |

**NOTATION**

| | | | |
|---|---|---|---|
| 11 | warp direction of woven ply | 23 | weft-through thickness direction of woven ply |
| 22 | weft direction of woven ply | 1t | warp tension direction of woven ply |
| 33 | through-thickness direction of woven ply | 1c | warp compression direction of woven ply |
| 12 | warp-weft direction of woven ply | 2t | weft tension direction of woven ply |
| 13 | warp-through thickness direction of woven ply | 2c | weft compression direction of woven ply |

## 1. Introduction

Woven composite laminates are now widely used in aerospace, transportation and renewable energy industries due to their impressive shear strength, impact resistance and fracture toughness compared to unidirectional laminates[1]. However, low-velocity impact (LVI) damage can significantly degrade the residual strength and fatigue life of woven composite laminates due to the lack of reinforcement in the thickness direction[2]. In recent decades, a large number of studies have reported that the post-impact fatigue (PIF) behaviours and failure mechanisms of composite laminates are affected by fibre type[3,4], laminate thickness[5], stacking sequence[6,7], impactor type[8], impact energy size[9,10] and environment condition[11], loading type[12,13].

The complex and interactive failure mechanisms in fatigue failure of woven composite laminates, along with the high computational cost, have impeded the development of predictive microscale physically-based fatigue damage models. Macroscale phenomenological models, such as residual strength model, are practical and efficient approaches to quantitatively predict the PIF life of composite laminates. Based on constant amplitude tension-tension (T-T) fatigue experiments of glass fibre laminates, Broutman and Sahu[14] firstly proposed linear degradation residual strength model, and predicted two-stage high-low (H-L) and low-high (L-H) fatigue life of undamaged laminates. It has been found that the model can provide more accurate life prediction results than Palmgren-Miner rule. Kang and Kim[15] and Koo et al.[16] developed fatigue life prediction models for post-impact composite laminates based on linear degradation residual strength model. Fatigue life prediction model was verified throuth constant amplitude T-T, two-stage H-L and L-H fatigue tests on impacted unidirectional and plain-weave carbon fibre laminates. Nevertheless, a large number of fatigue test results showed that the residual strength degradation curve of composite laminates follows a nonlinear rule. The linear degradation residual strength model cannot capture the nonlinear rule of residual strength[17,18], and neglects the significant load sequence effect for fatigue cumulative damage of composite laminates[19-22]. Therefore, it is necessary to develop a new fatigue life prediction model considering the nonlinear residual strength degradation rule and the effect of load sequence for composite laminates with LVI damage.

In order to decrease design cycle time and reduce experimental cost, virtual tests performed by finite element (FE) analysis are widely used to predict fatigue damage growth and residual fatigue life of post-impact composite structures. Mitrovic et al.[23] and Attia et al.[24] modelled interlaminar delamination by disconnecting the nodes of damaged region between the two sublaminates, and simulated fibre breakage and matrix cracking by the degradation of elastic moduli. A fatigue FE model of composite structures with the LVI damage was then established. Based on the fracture behaviours of composite materials, the damage growth behaviour of AS4/3501-6 laminates and fatigue life of post-impact carbon fibre reinforced polymer (CFRP) I-stringer stiffened panel was predicted. Recently, finite-element based progressive damage models have been widely used for investigating fatigue damage mechanisms and predicting fatigue life of composite structures. Shokrieh and Lessard[25,26] developed a fatigue failure criterion on the basis of the three-dimensional Hashin static failure criteria[27]. Residual strength and residual stiffness models of a unidirectional ply were used to characterise fatigue degradation properties of materials under cyclic loading, and sudden stiffness and strength degradation rules were adopted to represent material breakage. A fatigue progressive damage model was then built to predict constant amplitude fatigue life of composite laminates with three different stacking sequences and two-stage H-L and L-H fatigue life of quasi-isotropic laminates. The predictions correlated well with experimental results. Based on Puck's failure criteria[28], Passipoularidis et al.[29] proposed a stiffness degradation rule to represent the fatigue failure properties of materials and used residual strength model to characterise fatigue degradation performance of materials under cyclic loading. This was followed by a fatigue progressive damage model to predict variable amplitude fatigue life of glass fibre laminates. This model is able to capture fatigue behaviour considering load sequence effect. Wan et al.[30] presented a nonlinear fatigue-driven residual strength model considering stress ratio effect to describe fatigue degradation properties of laminates under cyclic loading, and applied with residual strength failure criterion and stiffness reduction rule. A fatigue progressive damage model then was used to assess fatigue life of helicopter composite tail structure under multipoint coordinated loading sequences.

In summary, a large number of experimental and theoretical research have been conducted to investigate fatigue behaviours and life prediction of composite laminates. Current theoretical and numerical works mainly focus on undamaged composite laminates. These studies on composite laminates with LVI damage are very limited due to the difficulties in characterising the LVI induced complex and interactive failure mechanisms and modelling LVI damage in fatigue FE model. Moreover, very few quantitative results can be found in literature to characterise the LVI effect on fatigue behaviours and life prediction of woven composite laminates under variable amplitude fatigue load. The main novel contributions herein are: (i) An original fatigue-driven residual strength model considering the effects of LVI damage and stress ratio was developed to characterise fatigue behaviour and to predict fatigue life of post-impact plain-weave composites. (ii) New fatigue failure criteria based on fatigue-driven residual strength concept were proposed to model fatigue damage initiation of plain-weave composites. (iii) A novel fatigue progressive damage model was established to simulate fatigue damage growth and predict fatigue life for plain-weave composite laminates with LVI damage. Our model opens new avenues for the efficient design of composite laminates against low-velocity impact and fatigue failure.

## 2. Fatigue-driven residual strength model considering the effects of LVI damage and stress ratio

A fatigue-driven residual strength model named $S$-$n$-$R$ model based on previous contribution[18] is proposed to characterise fatigue properties of pristine composite laminates at a specific stress ratio of $r_0$:

$$n = C\left(S - S_0\right)^p \left[X_0 - R\left(n\right)\right]^q \tag{1}$$

where $X_0$ is the static strength of composite laminates; $S$ is the maximum absolute value of fatigue stress; $R\left(n\right)$ is the fatigue residual strength after $n$ number of cycles; $n$ is the number of fatigue cycles; $S_0$ is the fatigue endurance limit of composite laminates; $C$, $p$ and $q$ are the material constants in fatigue residual strength model and are estimated from the constant amplitude fatigue test data $\left(S_i, n_i, R_i\right)$ by using best fitting.

It goes without saying that the static and fatigue strengths of post-impact composite laminates decrease with the increasing level of impact energy. If the impactor reaches the critical impact energy to penetrate the composite laminates, it can produce an open-hole damage with the similar diameter of impactor[16], so the post-impact static and fatigue strengths of composite laminates are assumed to be close to those of open-hole composite laminates. The following relations which are suitable to depict the characteristics of gradual reduction to a certain asymptotic line are presented to express the static and fatigue strengths of post-impact plain-weave laminates related to impact energy level:

$$X_0\left(\Pi\right) = \alpha_0 + \alpha_1 e^{-\alpha_2 \Pi} \tag{2}$$

$$S_0\left(\Pi\right) = \beta_0 + \beta_1 e^{-\beta_2 \Pi} \tag{3}$$

where $\Pi$ is the impact energy per unit thickness; $\alpha_0$, $\alpha_1$, $\alpha_2$, $\beta_0$, $\beta_1$ and $\beta_2$ are the model constants, and the model constants of $\alpha_0$, $\alpha_1$ and $\alpha_2$ are obtained from residual static strengths data $(\Pi_i, X_{0i})$ of undamaged and post-impact plain-weave laminates by means of best fitting method (see Appendix A).

Substituting Eq. (2) and Eq. (3) into Eq. (1), fatigue-driven residual strength model considering LVI damage effect is obtained as follow:

$$n = C\left[S - \left(\beta_0 + \beta_1 e^{-\beta_2 \Pi}\right)\right]^p \left[\left(\alpha_0 + \alpha_1 e^{-\alpha_2 \Pi}\right) - R(n)\right]^q \tag{4}$$

Then using the constant amplitude fatigue test results of undamaged and post-impact plain-weave laminate, the model constants of $C$, $p$, $q$, $\beta_0$, $\beta_1$ and $\beta_2$ are determined by the Least Squares Fitting method (see Appendix B).

Engineering composite structures often suffer from variable amplitude fatigue load under different stress ratios, so it is essential and desirable to account for the effect of stress ratio in the fatigue-driven residual strength model by using the modified Goodman curve as[21]:

$$\begin{cases} \dfrac{S_\mathrm{a}}{S_{-1}} + \dfrac{S_\mathrm{m}}{\sigma_\mathrm{TAI}} = 1 \quad \left(r^2 \le 1\right) \\ \dfrac{S_\mathrm{a}}{S_{-1}} + \dfrac{S_\mathrm{m}}{\sigma_\mathrm{CAI}} = 1 \quad \left(r^2 > 1\right) \end{cases} \tag{5}$$

where $S_\mathrm{a}$ and $S_\mathrm{m}$ are the amplitude and mean of nominal cyclic stress; $S_{-1}$ is the fatigue strength depicted with stress amplitude under fully reversed cyclic loading; $\sigma_\mathrm{TAI}=X_0$ is the tension after impact (TAI) strength of plain-weave laminate; $\sigma_\mathrm{CAI}=-X_0$ is the compression after impact (CAI) strength of plain-weave laminate.

Based on the definition of stress ratio and Eq. (5), it is possible to deduce the correction formulation from arbitrary stress ratio $r$ into specific stress ratio $r_0$ as follow:

$$S_{r_0} = \begin{cases} \dfrac{2\sigma_\mathrm{TAI}(1-r)S_\mathrm{max,r}}{(1-r_0)\left[2\sigma_\mathrm{TAI} - (1+r)S_\mathrm{max,r}\right] + (1+r_0)(1-r)S_\mathrm{max,r}} & \left(r^2 \le 1, r_0^2 \le 1\right) \\ \dfrac{-2\sigma_\mathrm{CAI}(r-1)r_0\left|S_\mathrm{min,r}\right|}{(r_0-1)\left[-2r\sigma_\mathrm{CAI} - (1+r)\left|S_\mathrm{min,r}\right|\right] - (1+r_0)(1-r)\left|S_\mathrm{min,r}\right|} & \left(r^2 > 1, r_0^2 > 1\right) \end{cases} \tag{6}$$

Substituting Eq. (2) and Eq. (6) into Eq. (4), fatigue-driven residual strength model considering the effects of LVI damage and stress ratio is determined as

$$\begin{cases} n = C\left\{\dfrac{2\left(\alpha_0+\alpha_1 e^{-\alpha_2 \Pi}\right)(1-r)S_{\text{max,r}}}{(1-r_0)\left[2\left(\alpha_0+\alpha_1 e^{-\alpha_2 \Pi}\right)-(1+r)S_{\text{max,r}}\right]+(1+r_0)(1-r)S_{\text{max,r}}}-\left(\beta_0+\beta_1 e^{-\beta_2 \Pi}\right)\right\}^{p}\left[\left(\alpha_0+\alpha_1 e^{-\alpha_2 \Pi}\right)-R(n)\right]^{q} \\ \left(r^2\le 1, r_0^2\le 1\right) \\ n = C\left\{\dfrac{2\left(\alpha_0+\alpha_1 e^{-\alpha_2 \Pi}\right)(r-1)r_0\left|S_{\text{min,r}}\right|}{(r_0-1)\left[2r\left(\alpha_0+\alpha_1 e^{-\alpha_2 \Pi}\right)-(1+r)\left|S_{\text{min,r}}\right|\right]-(1+r_0)(1-r)\left|S_{\text{min,r}}\right|}-\left(\beta_0+\beta_1 e^{-\beta_2 \Pi}\right)\right\}^{p}\left[\left(\alpha_0+\alpha_1 e^{-\alpha_2 \Pi}\right)-R(n)\right]^{q} \\ \left(r^2> 1, r_0^2> 1\right) \end{cases} \quad (7)$$

The nonlinear fatigue-driven residual strength model of Eq. (7) and fatigue residual strength failure criterion predict fatigue failure when the absolute fatigue stress exceeds the residual strength. This model can therefore be applicable to predict variable amplitude fatigue life of composite laminates for arbitrary LVI energy level and stress ratio. In the case of a given impact energy and stress ratio, fatigue-driven residual strength model of Eq. (7) can be reduced to the $S$-$n$-$R$ model. Then fatigue-driven residual strengths of the post-impact composite laminate at $n$ and $n+1$ loading cycles can be deduced as

$$\begin{cases} R(n) = X_0(\Pi)-\left\{nC^{-1}\left[S_{r_0}-S_0(\Pi)\right]^{-p}\right\}^{\frac{1}{q}} \\ R(n+1) = X_0(\Pi)-\left\{(n+1)C^{-1}\left[S_{r_0}-S_0(\Pi)\right]^{-p}\right\}^{\frac{1}{q}} \end{cases} \quad (8)$$

Taking transformation of Eq. (8) gives

$$R(n+1) = X_0(\Pi)-\left\{\left[X_0(\Pi)-R(n)\right]^{q}+C^{-1}\left[S_{r_0}-S_0(\Pi)\right]^{-p}\right\}^{\frac{1}{q}} \quad (9)$$

Eq. (9) is an iterative formula for residual strength of post-impact composite laminates under a given load sequence, which is a function of fatigue stress, loading cycles and impact energy. In order to characterise the difference in the TAI and CAI strengths of plain-weave composite laminate, the ratio between actual and maximum allowable strength degradation is adopted to adjust the residual strength of composite laminate when transition between tension-dominated and compression-dominated fatigue cycles occurs. According to Eq. (9), the fatigue-driven residual strength for composite laminate with LVI damage is calculated for each cycle. If fatigue residual strength failure criterion is triggered in current loading cycle, fatigue failure of composite laminate occurs and the predicted fatigue life is the total accumulative fatigue cycles.

## 3. Fatigue-driven failure criteria

Anisotropy nature of composites generally contributes to multiaxial stress state even under uniaxial fatigue loading (e.g., a cross-ply composite laminate), so it is important to determine multiaxial fatigue-driven residual strength. Residual strength degradation of a plain-weave ply under uniaxial fatigue loading conditions in warp and weft fibre direction and under in-plane shear fatigue loading condition generally occurs for plain-weave composite laminate under repeated fatigue cyclic loading, which can also be characterised by Eq. (1). Similarly, the modified Goodman curve is used to account for the effect of stress ratio, and the multiaxial fatigue-driven residual strength model considering stress ratio effect for plain-weave composite materials is obtained as,

$$\begin{cases} R_{it}(n) = X_{it} - \left\{ \left[ X_{it} - R_{it}(n-1) \right]^{q_{it}} + C_{it}^{-1} \left( S_{r_0} - S_{0,it} \right)^{-p_{it}} \right\}^{\frac{1}{q_{it}}} & (i=1,2) \\ R_{ic}(n) = X_{ic} - \left\{ \left[ X_{ic} - R_{ic}(n-1) \right]^{q_{ic}} + C_{ic}^{-1} \left( S_{r_0} - \sigma_{0,ic} \right)^{-p_{ic}} \right\}^{\frac{1}{q_{ic}}} & (i=1,2) \\ R_{12}(n) = X_{12} - \left\{ \left[ X_{12} - R_{12}(n-1) \right]^{q_{12}} + C_{12}^{-1} \left( S_{12} - S_{0,12} \right)^{-p_{12}} \right\}^{\frac{1}{q_{12}}} \end{cases} \tag{10}$$

with

$$S_{r_0} = \begin{cases} \dfrac{2X_{it}(1-r)S_{it\max,r}}{(1-r_0)\left[2X_{it} - (1+r)S_{it\max,r}\right] + (1+r_0)(1-r)S_{it\max,r}} & (i=1,2),\left(r_0^2 \le 1, r^2 \le 1\right) \\ \dfrac{2X_{ic}(r-1)r_0\left|S_{ic\min,r}\right|}{(r_0-1)\left[2X_{ic}r - (1+r)\left|S_{ic\min,r}\right|\right] - (1+r_0)(1-r)\left|S_{ic\min,r}\right|} & (i=1,2),\left(r_0^2 > 1, r^2 > 1\right) \end{cases} \tag{11}$$

$$S_{12} = \begin{cases} \dfrac{2X_{12}(1-r)S_{12\max,r}}{(1-r_0)\left[2X_{12} - (1+r)S_{12\max,r}\right] + (1+r_0)(1-r)S_{12\max,r}} & \left(r_0^2 \le 1, r^2 \le 1\right) \\ \dfrac{2X_{12}(r-1)r_0\left|S_{12\min,r}\right|}{(r_0-1)\left[2X_{12}r - (1+r)\left|S_{12\min,r}\right|\right] - (1+r_0)(1-r)\left|S_{12\min,r}\right|} & \left(r_0^2 > 1, r^2 > 1\right) \end{cases} \tag{12}$$

where $X_{1t}$, $X_{1c}$, $X_{2t}$, $X_{2c}$ and $X_{12}$ are the composite materials' warp tension and compression strengths, weft tension and compression strengths and in-plane shear strength, respectively; $C_{it}$, $p_{it}$, $q_{it}$, $S_{0,it}$, $C_{ic}$, $p_{ic}$, $q_{ic}$, $S_{0,ic}$, $C_{12}$, $p_{12}$, $q_{12}$ and $S_{0,12}\,(i=1,2)$ are the multiaxial fatigue-driven residual strength model constants, and those can be determined by the method shown in Appendix C. Although the Hashin failure criteria[27] have been frequently used for predicting failure responses of composites under static loading, but not in predicting fatigue failure mode because it neglects gradual strength degradation under cyclic loading. For this reason, the material strength in Hashin's criteria is replaced by multiaxial fatigue-driven residual strength (or Eq. (10)) to derive fatigue-driven Hashin failure criteria as follow:

Warp fibre tension fatigue failure ( $\sigma_{11} \geq 0$ ):

$$\left(\frac{\sigma_{11}}{X_{1t}-\left[X_{1t}-R_{1t}(n-1)\right]f_{1t}}\right)^2+\left(\frac{\sigma_{12}}{X_{12}-\left[X_{12}-R_{12}(n-1)\right]f_{12}}\right)^2 \geq 1 \tag{13}$$

Warp fibre compression fatigue failure ( $\sigma_{11} < 0$ ):

$$\left(\frac{\sigma_{11}}{X_{1c}-\left[X_{1c}-R_{1c}(n-1)\right]f_{1c}}\right)^2 \geq 1 \tag{14}$$

Weft fibre tension fatigue failure ( $\sigma_{22} \geq 0$ ):

$$\left(\frac{\sigma_{22}}{X_{2t}-\left[X_{2t}-R_{2t}(n-1)\right]f_{2t}}\right)^2+\left(\frac{\sigma_{12}}{X_{12}-\left[X_{12}-R_{12}(n-1)\right]f_{12}}\right)^2 \geq 1 \tag{15}$$

Weft fibre compression fatigue failure ( $\sigma_{22} < 0$ ):

$$\left(\frac{\sigma_{22}}{X_{2c}-\left[X_{2c}-R_{2c}(n-1)\right]f_{2c}}\right)^2 \geq 1 \tag{16}$$

In-plane shear fatigue failure ( $\sigma_{11} < 0$ ):

$$\left(\frac{\sigma_{11}}{X_{1c}-\left[X_{1c}-R_{1c}(n-1)\right]f_{1c}}\right)^2+\left(\frac{\sigma_{12}}{X_{12}-\left[X_{12}-R_{12}(n-1)\right]f_{12}}\right)^2 \geq 1 \tag{17}$$

where

$$\begin{cases} f_{it}=\left[\dfrac{\left[X_{it}-R_{it}(n-1)\right]^{-q_{it}}}{C_{it}\left(S_{it\max}-S_{0,it}\right)^{p_{it}}}+1\right]^{\frac{1}{q_{it}}} & (i=1,2) \\ f_{ic}=\left[\dfrac{\left[X_{ic}-R_{ic}(n-1)\right]^{-q_{ic}}}{C_{ic}\left(\left|S_{ic\min}\right|-S_{0,ic}\right)^{p_{ic}}}+1\right]^{\frac{1}{q_{ic}}} & (i=1,2) \\ f_{12}=\left[\dfrac{\left[X_{12}-R_{12}(n-1)\right]^{-q_{12}}}{C_{12}\left(S_{12}-S_{0,12}\right)^{p_{12}}}+1\right]^{\frac{1}{q_{12}}} & \end{cases} \tag{18}$$

# 4. Experimental procedures

## 4.1. Materials and specimens

The test specimens were made of plain-weave glass fibre reinforced polymer 3238A/EW250F, and materials' mechanical properties are presented in Table 1. Laminate specimens with the stacking sequences of $[(45/\text{-}45)/(0/90)]_{3s}$ and $[(45/\text{-}45)/(0/90)]_{2s}$ have same nominal ply thickness of 0.30 mm.

Fig. 1 shows the geometry and schematic of the specimen. The laminate plates were fabricated by a vacuum bag process under high temperature of 130 °C and pressure of 0.5 MPa, and then were cut by a water jet.

In order to obtain post-impact laminate specimens, the LVI tests of pristine plain-weave composite laminates were carried out on a HIT230F drop hammer impact tester at room temperature and moisture (see Fig. 2(a)), and the drop-weight has a mass of 5.61 kg with a hemispherical steel nose of 16 mm diameter. According to ASTM D7136[31], the specimen was firstly placed on the rigid base with a rectangular open hole of 75 mm × 125 mm size, and the specimen centre was then adjusted to align the impact location, after this, four corners of specimen were clamped by the support fixtures. During tests, all specimens were struck by the free drop-weight at a specific level of impact energy, which was controlled by adjusting the height of drop-weight from the rigid fixture base. The impact characteristics including contact force versus time and contact force versus deflection curves were automatically recorded by the test system. For specimens of $[(45/-45)/(0/90)]_{3s}$, the LVI tests of 9.79 J/mm, 12.19 J/mm and 15.07 J/mm were implemented to obtain three different extent impact damage, and the total number of specimens is 130. The specimens of $[(45/-45)/(0/90)]_{2s}$ were impacted with the energy level of 12.53 J/mm, and the number of specimens is 10.

### 4.2. Fatigue tests

According to ASTM D3479[32]and ASTM D7137[33], all constant amplitude and multi-step fatigue tests were carried out on the QBS-100 KN servo-hydraulic tester under the sinusoidal waveform loading with a frequency of 10 Hz at room temperature and moisture (see Figs. 2(b) and 2(c)). Firstly, the constant-amplitude fatigue tests for undamaged and post-impact laminate specimens of $[(45/-45)/(0/90)]_{3s}$ were conducted under tension-tension (T-T) and compression-compression (C-C) cyclic loading at the stress ratios of 0.05 and 10, respectively. The anti-buckling fixture was used to avoid the global buckling of the specimen under C-C loading[34]. For each specimen type, at least four groups of fatigue tests were conducted under different stress levels to achieve four target fatigue lives. Each group contained at least three specimens to ensure the reliability of the test data. The specimen which did not fail at the targeted fatigue life under the chosen fatigue stress level was loaded to failure under static tensile or compressive load to obtain the residual fatigue strength.

Moreover, the spectrum tests for 12.53 J/mm impacted $[(45/-45)/(0/90)]_{2s}$ specimens and 15.07 J/mm impacted $[(45/-45)/(0/90)]_{3s}$ specimens were implemented for validating the proposed model. Fig. 3 shows the load history of tension-dominated two-stage high-low (H-L) and low-high (L-H) fatigue loads and repeat high-low-high (H-L-H) fatigue load for $[(45/-45)/(0/90)]_{2s}$ laminates after 12.53 J/mm impact, and Fig. 4 illustrates the load history of compression-dominated two-stage H-L and L-H fatigue loads and repeat H-L-H fatigue load for 15.07 J/mm impacted $[(45/-45)/(0/90)]_{3s}$ laminates. The '$s$' is the maximum tensile stress when the specimen is subjected to T-T fatigue stress cycles at the stress ratio of 0.05. It is the minimum compressive stress when the specimen is subjected to C-C fatigue stress cycles at the stress ratio of 10. Similar to the constant amplitude fatigue test, at least three specimens were employed for each block-loading fatigue test to ensure the reliability of the test data.

## 5. Experimental results and discussion

### 5.1. Constant amplitude fatigue tests

Based on constant amplitude fatigue results from section 4.2, the Least Squares Fitting method (Appendix B) is used to obtain nonlinear fatigue-driven residual strength model considering the effects of LVI damage and stress ratio for 3238A/EW250F composite laminate as shown in Eq. (D.1). At experimental stress ratio $r$ of 0.05 and10, when residual strength $R$ in Eq. (D.1) equals to fatigue stress $S$, aforementioned fatigue-driven residual strength model reduces to Eq. (D.2).

The comparison between the $S$-$N$ curves of the 3238A/EW250F laminates characterised by Eq. (D.2) and test results is shown in Fig. 5. It is worth noting that the data points labelled with arrows and numbers represent the specimens which did not fail at the targeted fatigue life. Fig. 6 shows the comparison between the $S$-$n$-$R$ surfaces of 3238A/EW250F laminates at experimental stress ratio represented by Eq. (D.1). To assess the effect of the LVI damage on 3238A/EW250F laminates under constant amplitude T-T and C-C loading, the fatigue strength ratio is defined as the ratio of fatigue strength of post-impact laminate to fatigue strength of pristine laminate at one million fatigue cycles, and is shown in Fig. 7. From Figs. 5-7, three major findings are as follows:

i) Based on nonlinear fatigue-driven residual strength model considering the effects of LVI damage and stress ratio, both derived $S$-$N$ curve and $S$-$n$-$R$ surface correlate well with constant amplitude fatigue test results of post-impact 3238A/EW250F laminates, indicating that it is effective for developed fatigue-driven residual strength model to characterise fatigue properties of 3238A/EW250F laminates with LVI damage.

ii) The $S$-$N$ curves of post-impact woven laminates are flatter than those of pristine woven laminates. The LVI damage magnitude of specimen and stress concentration in damage region increase with the increasing of the impact energy level. This leads to a gradual reduction in fatigue strength of post-impact plain-weave laminates under constant amplitude T-T and C-C fatigue load, showing the expected degradation effect of LVI damage.

iii) Fatigue strength ratio curve under C-C fatigue load is below that under T-T fatigue load, indicating that the degradation effect of LVI damage on fatigue strengths of the 3238A/EW250F laminate is larger under C-C fatigue load than T-T fatigue load. This is due to the fact that the plain-weave laminate with LVI damage is prone to local buckling under C-C fatigue load even if anti-buckling device was employed, resulting in easier fatigue fracture under C-C fatigue load than under T-T fatigue load[35].

### 5.2. Multi-step fatigue tests

Table 2 presents fatigue test results of 3238A/EW250F laminates with 12.53 J/mm and 15.07 J/mm impact damage under multi-step fatigue load shown in Figs. 3 and 4. The cumulative fatigue damage under variable amplitude fatigue load is calculated by the Palmgren–Miner linear damage model as

$$D=\sum_{i=1}^{k}\frac{n_i}{N_i} \tag{19}$$

where $n_i$ and $N_i$ are the cycle number and constant amplitude fatigue life under fatigue stress of $S_i$, respectively. The used fraction of life is defined as $n_i / N_i$. Although the Palmgren–Miner linear damage model (or Eq. (19)) cannot effectively reflect the load sequence effect, it can be used to calculate the cumulative fatigue damage value under variable amplitude fatigue load, which is related to the load sequence. And the calculated results of cumulative fatigue damage can still probably mark and illustrate the load sequence effect. According to Eq. (D.1), the $S$-$n$-$R$ model for 12.53 J/mm and 15.07 J/mm impacted plain-weave laminates at experimental stress ratio are deduced as shown in Eq. (D.3).

Eq. (D.3) reduces to the $S$-$N$ curve as residual strength $R$ equals to fatigue stress $S$. Based on the curve and Eq. (19), the cumulative fatigue damage values $D$ of post-impact plain-weave laminates under multi-step fatigue load are calculated as listed in Table 2. Moreover, according to the $S$-$n$-$R$ surface of Eq. (D.3a) and fatigue-driven residual strength iterative formula (see Eq. (9)), strength degradation behaviour for 12.53 J/mm impacted plain-weave laminates under tension-dominated two-stage fatigue load is described in Fig. 8. From Table 2 and Fig. 8, three main findings can be obtained as follows:

i) The accumulation fatigue damage for post-impact 3238A/EW250F laminates under tension-dominated and compression-dominated two-stage H-L and L-H load sequences do not follow a linear accumulation rule. There is no effect of load sequence if the Palmgren–Miner damage value $D$ equals to 1.0 under the ideal fatigue load condition. The damage value $D$ of 3238A/EW250F laminates with LVI damage is less than 1.0 under L-H load sequence means that more damage occurs under L-H load sequence compared to ideal state, so fatigue failure occurs prematurely. On the contrary, if the damage value $D$ of post-impact 3238A/EW250F laminates is greater than 1.0 under H-L load sequence, it means that less damage occurs under H-L load sequence compared to ideal condition, so fatigue failure delays. The above results demonstrates that there is a strong load sequence effect on fatigue damage of post-impact plain-weave composite laminates, and the L-H load sequence is more damaging than the H-L sequence, which is similar with the findings in literatures[22].

ii) Post-impact 3238A/EW250F laminates under two-stage H-L and L-H sequences fail at different strength levels, and the maximum allowable strength degradation under L-H sequence is smaller than that under H-L sequence when laminate failure happens (see Fig. 8(a)). As shown in Fig. 8(b), under two-stage H-L sequence, residual strength of composite laminate reduces to the value of $R\left(n_H / N_H\right)$ at the used fraction life $n_H / N_H$ in the first stage. In the second stage, high fatigue stress changes to the low fatigue stress, and residual strength of composite laminate will degrade from the value of $R\left(n_H / N_H\right)$ to a final residual strength. Owing to the residual strength values of $R\left(n_L / N_L\right)$ and $R\left(n_H / N_H\right)$ are equal and the second stage fatigue stress is at low fatigue stress loading, the residual strength degradation curve from point A to B is parallel and equal to that curve from point A' to B'. This leads to a longer cumulative used fraction of life under H-L sequence than the case under pure low fatigue stress condition (see symbols $D_{\text{H-L}}$ and $D_{\text{L}}$ in Fig. 8(b)). Considering the microstructural failure mechanisms, the dominant failure mode in high fatigue stress is fibre-dominated damage while matrix-dominated damage is often observed in the low fatigue stress level[9,36]. In the high fatigue stress regime of the H-L sequence, fibre breakage is the dominant failure mode whilst the development of matrix-dominated damage is impeded. With less matrix damage at the beginning of the low fatigue stress stage, the fatigue tolerance of composite laminates increases, thereby extending the total fatigue life[21]. Similarly, the cumulative used fraction of life under L-H sequence to be smaller than that under pure high fatigue stress can also be explained. The above explanation demonstrates that the improved fatigue-driven residual strength model in this paper can effectively reflect the load sequence effect.

iii) Compared to tension-dominated and compression-dominated two-stage H-L and L-H sequences, the cumulative fatigue damage values of post-impact plain-weave composite laminates under repeated H-L-H fatigue loading including tension and compression fatigue stress are smallest (see Table 2), showing the most severe cumulative fatigue damage. It indicates that the interactions between fatigue stresses with several stress ratios and stress levels significantly shorten fatigue life of plain-weave composites with LVI damage. This is similar to the findings in literature[37] that frequent transitions of fatigue stress will cause substantial damage. As shown in Figs. 3 and 4, the transition between the tensile and compressive stress under repeated H-L-H fatigue loading forms a large stress amplitude loading cycle with nearly twice the stress amplitude of the high fatigue stress. This kind of large transition loading cycle may trigger stronger failure mode interactions and hence can significantly shorten fatigue life of plain-weave composite laminates[21].

### 5.3. Fractographic analysis

To understand the fatigue failure mechanisms of post-impact 3238A/EW250F laminates, fatigue fracture specimens under constant amplitude and two-stage fatigue loads are first observed in naked eyes (shown in Fig. 9). For the fatigue specimen under constant amplitude T-T loading shown in Fig. 9(c), four typical positions are examined in a scanning electron microscope (JEOL JSM-6010) (shown in Figs. 10(a)-10(d)). Figs. 11(a)-11(c) show the observations of three characterised sites in the specimen under constant amplitude C-C fatigue load shown in Figs. 9(e) and 9(f). In addition, Fig.12 describes fractographic observations of fatigue specimens under tension-dominated and compression dominated two-stage H-L and L-H fatigue loads. From experimental observations (shown in Figs. 9-12), results are summarized as follows:

i) Fatigue fracture appears on the stress concentration sites around the dent of front surface of post-impact laminates (see Fig. 9(a) and 9(d)), while two typical macroscopic failures patterns randomly emerge on the back surface. This may be related to the randomness of the stress concentration at the border of damaged area. One is the specimen failed at two top (or bottom) crack tips of the X-type damage (see Figs. 9(b) and 9(e)), and the other is the specimen fractured at the top right and bottom left crack tips of the X-type damage (see Figs. 9(c) and 9(f)).

ii) Under constant amplitude T-T fatigue load, the transverse cracks induced by the LVI propagate into the laminates, and their density increases with the increasing number of loading cycles. After crack density reaches saturation, those cracks coalesce and further develop into interface debonding and local delamination, and impact-induced delamination also propagate during this period (see Fig. 10(a)). Multiple delaminations are prone to cause the separation of the layers. In this way, the layers may fail in different planes due to the different strengths of each layer, leading to the brooming failure mode[38]. The delamination mainly propagates perpendicular to the loading direction to both sides of the specimen and rarely propagates along the loading direction (see Fig. 10(b)). Delamination and fibre tensile failure successively initiate and propagate to both sides of the specimen where the brooming failure mode is observed (see Figs. 10(c) and 10(d)), that is, there are a large number of small delaminations and fibre breakages as shown in Figs. 10(e) and 10(f).

iii) Under constant amplitude C-C fatigue load, significantly different fatigue failures are observed compared to constant amplitude T-T fatigue load. Impact-induced delamination promotes the buckling behaviour of interior layers. As the fatigue cycles increase, the delamination splits and propagates in the shear planes, leading to the wedge splitting failure mode. There is a large delamination in the middle of the laminate, and some front surface's layers are crushed on the fracture plane due to the buckling behaviour of the specimen towards the dent (see Fig. 11(a)). With the increase in the number of fatigue cycles, delamination and buckling continue to propagate, and wedge splitting propagates irregularly in the width direction from LVI damage area to both sides of the specimen. Finally, the wedge splitting failure mode or mixed failure mode of through-thickness shear and delamination is formed on both sides of the specimen, which is a typical compression failure mode for plain-weave laminates[39]. Specifically, wedge splitting failure mode is characterised by two different shear fractures propagating in the thickness direction and, in the encounter of these shear fractures a wedge is formed and a large delamination propagates (see Fig. 11(b)). The mixed failure mode of through-thickness shear and delamination is that the damages propagate at an angle to the thickness direction, accompanied by some delaminations (see Fig. 11(c)).

iv) Brooming failure mode also emerges on both sides of specimen under tension-dominated two-stage H-L and L-H sequences (see Figs. 12(a) and 12(b)), which is similar to the fracture morphology (see Fig. 10(c)) under constant amplitude T-T fatigue load. The slight difference is that delamination propagates more widely in the loading direction under L-H sequence than that under H-L sequence. Under compression-dominated two-stage H-L sequence, the fatigue failure specimen tends to form the through-thickness shear failure, accompanied by multiple delaminations (see Fig. 12(c)). Differently, wedge splitting failure mode is often observed under compression-dominated two-stage L-H sequence (see Fig. 12(d)).

## 6. Progressive damage analysis

### 6.1. LVI progressive damage analysis

The LVI progressive damage analysis of plain-weave composite laminate is a basis for predicting the PIF life, and this process aims to obtain reliable LVI damage of the plain-weave laminate. The plain-weave laminate shown in Fig. 1 is modelled, and the local coordinates are set up to ensure each woven layer with correct 3D orientation, namely, to keep three axial directions x, y and z of the coordinate system consistent with the warp, weft and through-thickness directions (or three normal-stress directions) for each woven layer (see Fig. 13). The 3D element C3D8R are employed to model each layer of plain-weave laminate, and the element number of each layer is 5625. The element C3D8R is also used to model rigid base of 300 mm × 300 mm × 10 mm size with a rectangular open hole, and the number of elements is 6738. Moreover, rigid element R3D4 is used to model the hemispherical impactor, and the number of elements is 1726. In addition, the cohesive element COH3D8 is embedded in the middle of adjacent layers with a thickness of 0.01 mm, and the element number of each cohesive ply is 5625. Thus, a 3D FE model of plain-weave composite laminate (shown in Fig. 13) is generated by ABAQUS code to simulate the LVI response in association with relevant mechanical properties listed in Table 1 and material properties of cohesive element listed in Table 3. The boundary conditions in 3D FE model are defined as the clamped constraints on interior and exterior sides of the rigid base and four sides of the laminate model. The inertial mass and impact velocity are loaded at the tip of impactor for modelling the specific impact energy (see Fig. 13(b)).

In the LVI progressive damage simulation, the failure mode of interlaminar delamination is characterised by the ABAQUS built-in cohesive zone model, in which maximum stress criterion and B-K rule are generally used for identifying delamination initiation and evolution[34]. Other failure modes identification and relevant material property degradation are realised in progressive damage algorithm, which is written in the VUMAT subroutine. Schematic flowchart for progressive damage analysis is shown in Fig. 14(a). For each iterative calculation, after stress state of elements is calculated, the failure of element is identified by Olmedo failure criteria[40]. Similar to continuum damage mechanics, a sudden stiffness degradation model (shown in Table 4)[25,30] is used for predicting impact and fatigue damage. This rule directly defines the final degraded stiffnesses corresponding to different failure modes of Hashin's criteria, and the parameters used in this rule are calibrated against the LVI experimental results. If reaching the failure criteria, the sudden stiffness degradation rule is used to degrade the stiffness of elements, and the stress state for each element is then updated for further identifying the failure by using the LVI failure criteria until the end of impact time.

The predicted contact force versus time and deflection curves for $[(45/-45)/(0/90)]_{3s}$ laminates subjected to 12.19 J/mm impact are shown in Fig. 15. From Fig. 15, it is apparent that the LVI simulation results are in good agreement with the experiments. The predicted maximum peak force and peak deflection are 7914 N and 9.97 mm, and the experimental values of those are 7776 N and 9.89 mm, so the relative deviations are within 2%. Consequently, this confirms that the LVI progressive damage analysis can accurately predict the LVI behaviours for plain-weave composite laminates, and reliably afford LVI damage.

### 6.2. PIF progressive damage analysis

In the PIF progressive damage analysis of plain-weave composite laminates, the FE model with LVI damage is firstly imported. Based on the LVI results, the failed elements are identified to obtain post-impact FE model, where each intralaminar and interlaminar layer contains 5625 C3D8R and COH3D8 elements, respectively. The impact-induced delamination is simulated by means of deleting the failed COH3D8 elements. It is worth noting that the initiation and evolution of interlaminar delamination for undamaged COH3D8 elements is still characterised by aforementioned cohesive zone model. Moreover, failed intralaminar elements inherit degradation material properties, while intact C3D8R elements are given mechanical properties as shown in Table 1 and multiaxial fatigue-driven residual strength properties (see Table 5) as characterised by Eq. (10). Material constants (listed in Table 5) for fatigue-induced residual strength model are estimated from axial constant amplitude fatigue tests[41] of plain-weave 3238A/EW250F laminates $[(0/90)]_9$ and $[(45/-45)]_9$ by using the fitting method presented in Appendix C. The left side of laminate is fully fixed and fatigue load is applied on the right surface in pressure. The top and bottom sides of laminate are constrained to model the boundary constraint of anti-buckling device in the compression-dominated fatigue cycles (see Fig. 16). Noticeably, each loading cycle is modelled into the quasi-static loading with the same magnitude as the maximum absolute value of the loading cycle.

Schematic flowchart for the PIF progressive damage analysis is shown in Fig. 14(b). The PIF progressive damage algorithm is written in the VUMAT subroutine, in which layer's material property degradation and element failure identification are realised. Multi-stage fatigue load history (shown in Figs. 3 and 4) is applied to the FE model of plain-weave laminates. After stress state analysis of FE model, fatigue residual strength of intralaminar element is gradually degraded by using Eq. (10) and developed fatigue failure criteria (or Eqs. (13) to (17)) are updated. The failure of intralaminar element is then identified by fatigue failure criteria. If failure happens, the stiffness of failed element is degraded to nearly zero according to sudden stiffness reduction rule and the FE stress analysis is thereafter re-executed.

The damage propagation in 15.07 J/mm post-impact plain-weave laminates under compression-dominated two-stage H-L fatigue load is illustrated in Fig. 17, in which impact-induced damage elements are presented in white colour and failed elements during fatigue process are marked with red colour. From Fig. 17, it can be seen that when first-stage of the H-L sequence runs out, that is, 18500 fatigue cycles, the elements of warp fibre failure are only found at (0/90) layers, some elements of in-plane shear failure occur at each layer of (±45), and there are some interlaminar delamination failures between each layer (see Fig. 17(a)), which is in good agreement with the experimental observation. Finally, fatigue failure happens at 177232 cycles under the second-stage of the H-L sequence, at this time, warp and weft fibre failure, in-plane shear failure and delamination in the composite laminate all propagate perpendicular to the loading direction to both sides of the laminate (see Fig. 17(b)). Noticeably, the 5th, 7th and 8th interlaminar delaminations are greater than other delaminations, which is consistent with the observation of multiple larger delaminations found in the fatigue fracture morphology (see Fig. 12(c)). Overall, the predicted results correlate well with the experimental observations.

The multi-stage fatigue lives predicted by the PIF progressive damage model are shown in Table 2. In order to compare the prediction accuracy of different fatigue life models, Table 2 also lists fatigue life results predicted by the Palmgren-Miner rule and nonlinear fatigue-driven residual strength model considering the effects of LVI damage and stress ratio. By utilizing the Palmgren-Miner rule, the final fatigue life of composite laminates under spectrum loads is determined when the cumulative Miner damage value equals one. The life prediction procedure of fatigue-driven residual strength model developed in this paper has been given in Section 2. From Table 2, the maximum relative deviations between fatigue life predictions and experiments using Palmgren–Miner rule, fatigue-driven residual strength model and PIF progressive damage model are 113%, 72% and 17%, respectively. This demonstrates fatigue-driven residual strength model considering the effects of LVI damage and stress ratio and PIF progressive damage model can provide more accurate prediction results than Palmgren–Miner cumulative damage model, owing to both models reflecting the load sequence effect. Theoretically, because of the PIF fatigue progressive damage model using constant amplitude fatigue properties of undamaged composite materials to predict the spectrum life of laminates with LVI damage, this model can be used for the variable amplitude life prediction of laminates with any stacking sequence and geometric configuration.

Needless to say, the post-impact fatigue (PIF) progressive damage model is based on several fatigue test data for a single material system. Constant amplitude tension-tension (T-T) and compression-compression (C-C) fatigue properties of the plain-weave ply under uniaxial fatigue loading conditions in warp and weft fibre directions and under in-plane shear fatigue loading condition need to be obtained before applying the PIF progressive damage model. These fatigue properties are all obtained from composite laminates with unidirectional stacking sequence (eg. $[(0/90)]_9$ and $[(45/-45)]_9$), and the measured ply-level properties can be used as the parameters for the material constitutive laws. The generalized constitutive model is then defined for each finite element that can consider multiaxial stress state. Therefore, the PIF progressive damage model is applicable for complicate fatigue loading conditions, including bending conditions.

The RVE based multi-scale FE modeling method is a new avenue to address this problem, and the significant advantages of multi-scale FE modeling based on the fatigue properties of fibre, matrix and fibre-matrix interface are enough to chararterise complex woven structure (e.g., three-dimensional braided composites), and can save experimental cost and time without fatigue tests in several loading conditions. However, several challenges limit the use scope of this approach. First of all, it is hard to simulate the fatigue properties of laminates by using the RVE model because of the statistic character of fatigue data and size effect of specimens. Meantime, the fatigue properties of the fibre, matrix and fibre-matrix interface are very challenging to measure by the existing test set-ups. Furthermore, post-impact plain-weave laminate has complex and interactive failure mechanisms. The RVE model may not be capable to depict all the failure mechanisms, so the numerical errors between predictions and experiments could be transferred from the RVE model to ply-level model, to laminate-level model, and structure-level model, thereby making the final prediction not acceptable in practice. In addition, due to complex structural features of woven composite structures, multi-scale FE modeling is rather complex and the computational cost is rather huge at present, which limits its scope of use.

## 7. Conclusions

This paper presents an experimental and numerical study on fatigue life of post-impact composite laminate specimen made of plain-weave glass fibre lamiantes 3238A/EW250F under various fatigue load sequence, including constant-amplitude T-T ( $r = 0.05$ ) and C-C ( $r = 10$ ) fatigue load, and two-stage H-L, L-H fatigue load, and repeated H-L-H fatigue load. The following conclusions can be drawn from this study:

i) Low-velocity impact damage has significant degradation effect on fatigue strength of plain-weave composite laminates. The degradation effect is greater under compressive fatigue load than tensile fatigue load.

ii) Fatigue damage of post-impact plain-weave composite laminates is significantly affected by load sequence. The tension-dominated and compression dominated L-H sequences are more damaging than tension-dominated and compression dominated H-L sequences, while repeat H-L-H loading including tensile and compressive stress can cause the highest fatigue damage.

iii) The damage mechanisms of plain-weave composite laminates with LVI damage under constant amplitude T-T and C-C fatigue load are significantly different. Fatigue specimens under T-T fatigue load show brooming failure mode, while those under C-C fatigue load present wedge splitting failure mode and mixed failure mode of through-thickness shear and delamination.

iv) Improved nonlinear fatigue-driven residual strength model considering the effects of LVI damage, stress ratio and load sequence and fatigue progressive damage model using new fatigue failure criteria have been developed to predict fatigue life of plain-weave composite laminates with LVI damage. The fatigue life predictions of post-impact plain-weave composite laminates by using those models correlates well with the experimental results, showing the effectiveness and accuracy of proposed fatigue life prediction models

## Acknowledgements

This project was supported by the National Natural Science Foundation of China (Grant No. 51875021) and the China Scholarship Council.

## Appendix A

Transforming and taking the natural logarithm form of the Eq. (2) gives

$$u=\phi_0+\phi_1 v \tag{A.1}$$

where $u=\ln\left(X_0-\alpha_0\right)$, $\phi_0=\ln\alpha_1$, $\phi_1=-\alpha_2$ and $v=\Pi$.

By means of the linear regression principle, following results can be obtained

$$\begin{cases} \phi_0=\bar{u}-\dfrac{L_{vu}}{L_{vv}}\bar{v} \\ \phi_1=\dfrac{L_{vu}}{L_{vv}} \\ R^2\left(\alpha_0\right)=\dfrac{L_{vu}^2}{L_{vv}L_{uu}} \end{cases} \tag{A.2}$$

where $R^2\left(\alpha_0\right)$ is square of the correlation coefficient, and

$$\begin{cases} \bar{v}=\dfrac{1}{m}\sum_{i=1}^{m}v_i \\ \bar{u}=\dfrac{1}{m}\sum_{i=1}^{m}u_i \\ L_{vv}=\sum_{i=1}^{m}\left(v_i-\bar{v}\right)^2 \\ L_{uu}=\sum_{i=1}^{m}\left(u_i-\bar{u}\right)^2 \\ L_{vu}=\sum_{i=1}^{m}\left(v_i-\bar{v}\right)\left(u_i-\bar{u}\right) \end{cases} \tag{A.3}$$

where $m$ is the number of specimens.

By means of the maximum value principle of $R^2\left(\alpha_0\right)$, it is possible to determine the parameter $\alpha_0$, then intermediate variables $\phi_0$ and $\phi_1$ can be obtained. Finally, unknown model constants $\alpha_1$ and $\alpha_2$ are determined as

$$\begin{cases} \alpha_1=e^{\phi_0} \\ \alpha_2=-\phi_1 \end{cases} \tag{A.4}$$

**Appendix B**

The natural logarithm form of the Eq. (4) can be written as

$$y = a_0 + a_1 x_1 + a_2 x_2 \tag{B.1}$$

with

$$\begin{cases} a_0 = \ln C \\ a_1 = p \\ a_2 = q \\ x_1 = \ln\left[ S - \left( \beta_0 + \beta_1 e^{-\beta_2 \Pi} \right) \right] \\ x_2 = \ln\left[ \left( \alpha_0 + \alpha_1 e^{-\alpha_2 \Pi} \right) - R(n) \right] \\ y = \ln n \end{cases} \tag{B.2}$$

According to the maximum likelihood principle, the intermediate variables $a_0$, $a_1$, $a_2$ and residual sum of squares $Q$ can be obtained

$$\begin{cases} a_0 = \bar{y} - a_1 \bar{x}_1 - a_2 \bar{x}_2 \\ a_1 = \dfrac{L_{12} L_{20} - L_{22} L_{10}}{L_{12} L_{21} - L_{11} L_{22}} \\ a_2 = \dfrac{L_{21} L_{10} - L_{11} L_{20}}{L_{12} L_{21} - L_{11} L_{22}} \\ Q = \sum_{i=1}^{m} \left( y_i - a_0 - a_1 x_1 - a_2 x_2 \right)^2 \end{cases} \tag{B.3}$$

with

$$
\begin{cases}
\bar{y} = \dfrac{1}{m}\sum_{i=1}^{m} y_i \\
\bar{x}_1 = \dfrac{1}{m}\sum_{i=1}^{m} x_{1i} \\
\bar{x}_2 = \dfrac{1}{m}\sum_{i=1}^{m} x_{2i} \\
L_{11} = \sum_{i=1}^{m}\left(x_{1i} - \bar{x}_1\right)^2 \\
L_{22} = \sum_{i=1}^{m}\left(x_{2i} - \bar{x}_2\right)^2 \\
L_{12} = L_{21} = \sum_{i=1}^{m}\left(x_{1i} - \bar{x}_1\right)\left(x_{2i} - \bar{x}_2\right) \\
L_{10} = \sum_{i=1}^{m}\left(x_{1i} - \bar{x}_1\right)\left(y_i - \bar{y}\right) \\
L_{20} = \sum_{i=1}^{m}\left(x_{2i} - \bar{x}_2\right)\left(y_i - \bar{y}\right)
\end{cases}
\tag{B.4}
$$

From Eq. (B.2) to Eq. (B.4), it can be seen that $x_1$, $L_{11}$, $L_{12}$ and $L_{10}$ are functions of the undetermined constants $\beta_0$, $\beta_1$ and $\beta_2$. So the parameters $a_0$, $a_1$, $a_2$ and $Q$ are also the functions with regard to $\beta_0$, $\beta_1$ and $\beta_2$. By means of the minimum value principle of $Q$, one has

$$
\begin{cases}
\dfrac{\partial Q}{\partial \beta_0} = 0 \\
\dfrac{\partial Q}{\partial \beta_1} = 0 \\
\dfrac{\partial Q}{\partial \beta_2} = 0
\end{cases}
\tag{B.5}
$$

By numerically solving Eq. (B.5), the solutions of $\beta_0$, $\beta_1$ and $\beta_2$ can be obtained and the unknown model constants $C$, $p$ and $q$ are then determined as

$$
\begin{cases}
C = e^{\left(\bar{y} - a_1\bar{x}_1 - a_2\bar{x}_2\right)} \\
p = \dfrac{L_{12}L_{20} - L_{22}L_{10}}{L_{12}L_{21} - L_{11}L_{22}} \\
q = \dfrac{L_{21}L_{10} - L_{11}L_{20}}{L_{12}L_{21} - L_{11}L_{22}}
\end{cases}
\tag{B.6}
$$

**Appendix C**

Transforming Eq. (10) gives

$$n = C\left(S - S_0\right)^p \left[X_0 - R(n)\right]^q \tag{C.1}$$

the natural logarithm form of Eq. (C.1) is same to the form of Eq. (B-1), except $x_1 = \ln\left(S - S_0\right)$, $x_2 = \ln\left[X_0 - R(n)\right]$, other terms is consistent. According to the maximum likelihood principle, the parameters $a_0$, $a_1$, $a_2$ and $Q$ are obtained as shown in Eqs. (B.3) and (B.4), and they are the function with regard to $S_0$. Again, based on the minimum value principle of $Q$, it is possible to have

$$\frac{\partial Q}{\partial S_0} = 0 \tag{C.2}$$

By numerically solving Eq. (C.2), the solution of $S_0$ can be obtained and the unknown model constants $C$, $p$ and $q$ are then determined as shown in Eq. (B.6).

**Appendix D**

$$\begin{cases} n = 1.96\times10^{17}\left\{\dfrac{\left(170.92+530.222e^{-0.107\Pi}\right)(1-r)S_{\max,r}}{0.95\left[\left(170.92+530.222e^{-0.107\Pi}\right)-(1+r)S_{\max,r}\right]+1.05(1-r)S_{\max,r}} \right. \\ \qquad \left. -\left(1.486+54.636e^{-0.144\Pi}\right)\right\}^{-7.5}\left[\left(85.460+265.111e^{-0.107\Pi}\right)-R(n)\right]^{0.97}, \left(r^2 \le 1\right) \\ n = 5\times10^{8}\left\{\dfrac{\left(2224.16+2487.56e^{-0.085\Pi}\right)(r-1)\left|S_{\min,r}\right|}{9\left[\left(222.416+248.756e^{-0.085\Pi}\right)r-(1+r)\left|S_{\min,r}\right|\right]-11(1-r)\left|S_{\min,r}\right|} \right. \\ \qquad \left. -\left(70.776+87.480e^{-0.204\Pi}\right)\right\}^{-4.54}\left[\left(111.208+124.378e^{-0.085\Pi}\right)-R(n)\right]^{0.50}, \left(r^2 > 1\right) \end{cases} \tag{D.1}$$

$$\begin{cases} N = 1.96\times10^{17}\left[S_{\max}-\left(1.49+54.64e^{-0.14\Pi}\right)\right]^{-7.50}\left[\left(85.46+265.11e^{-0.11\Pi}\right)-S_{\max}\right]^{0.97}, (r=0.05) \\ N = 5.00\times10^{8}\left[\left|S_{\min}\right|-\left(70.78+87.48e^{-0.20\Pi}\right)\right]^{-4.54}\left[\left(111.21+124.38e^{-0.09\Pi}\right)-\left|S_{\min}\right|\right]^{0.50}, (r=10) \end{cases} \tag{D.2}$$

$$\begin{cases} n = 1.96\times10^{17}\left(S_{\max}-10.95\right)^{-7.50}\left[152.27-R(n)\right]^{0.97}, (r=0.05) \\ n = 5.00\times10^{8}\left(\left|S_{\min}\right|-77.92\right)^{-4.54}\left[151.48-R(n)\right]^{0.50}, (r=10) \end{cases} \left(\Pi = 12.53\right) \tag{D.3a}$$

$$\begin{cases} n = 1.96\times10^{17}\left(S_{\max}-8.11\right)^{-7.50}\left[135.98-R(n)\right]^{0.97}, (r=0.05) \\ n = 5.00\times10^{8}\left(\left|S_{\min}\right|-75.07\right)^{-4.54}\left[143.25-R(n)\right]^{0.50}, (r=10) \end{cases} \left(\Pi = 15.07\right) \tag{D.3b}$$

145: 106110.

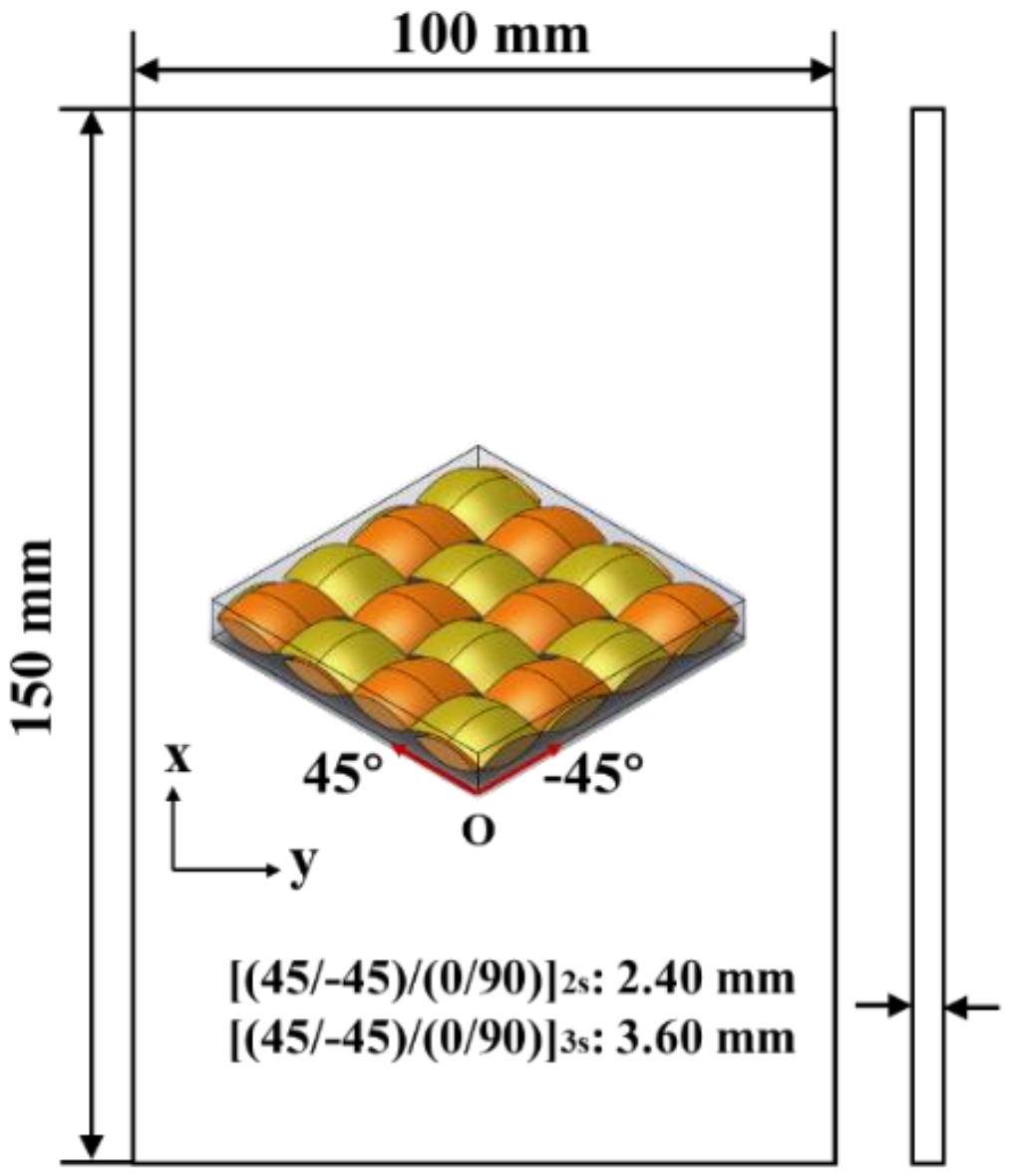

**Fig. 1** Geometric geometry and schematic of specimens.

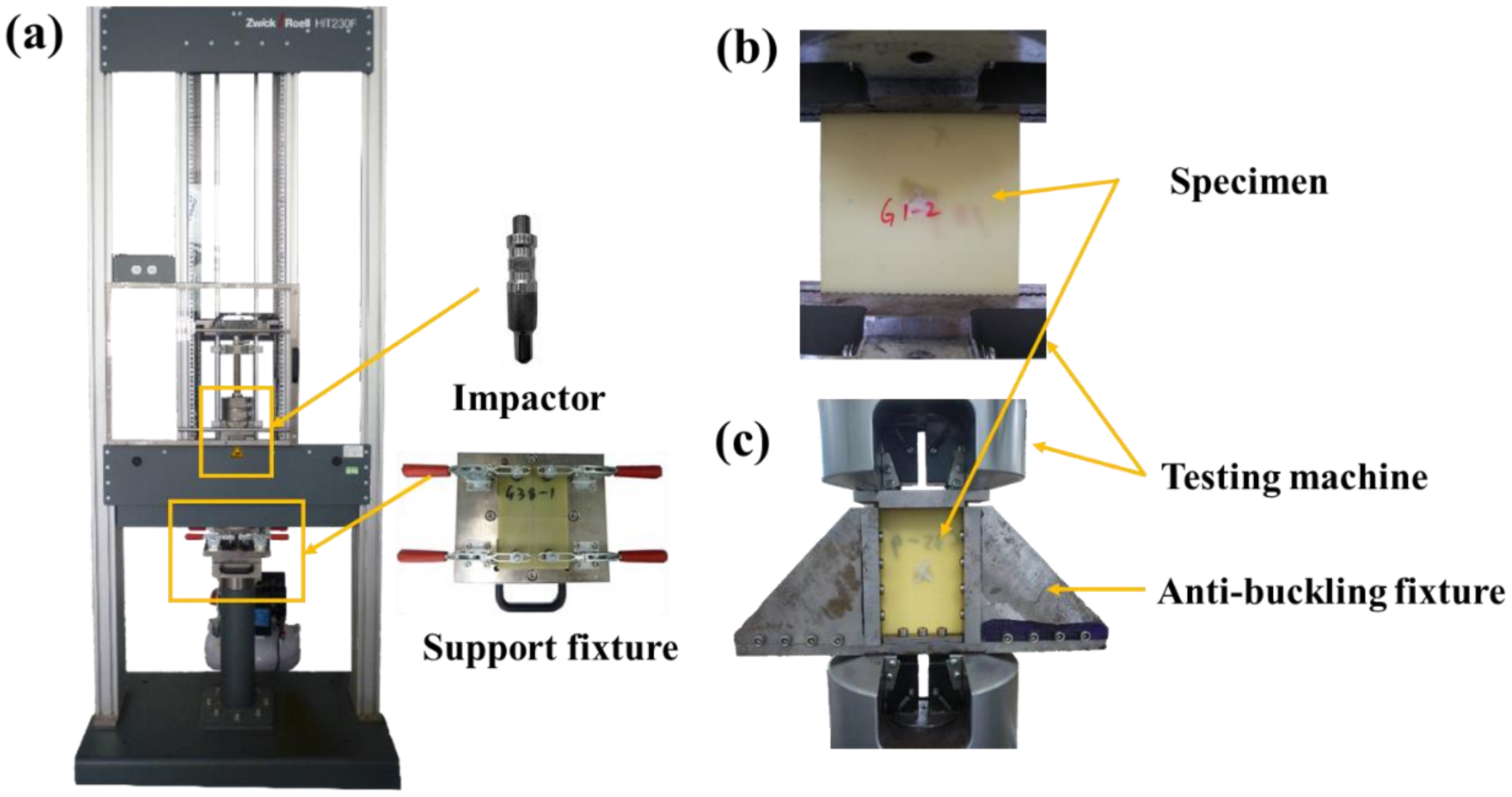

**Fig. 2** LVI and fatigue test setups: (a) HIT230F drop hammer impact tester, (b) tension-dominated fatigue test setup, (c) compression-dominated fatigue test setup.

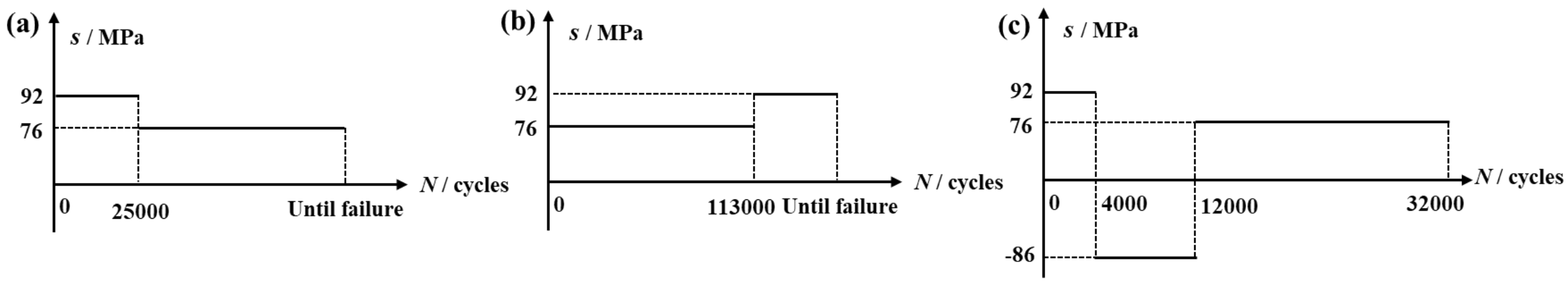

**Fig. 3** Load history of multi-step fatigue tests for $[(45/-45)/(0/90)]_{2s}$ laminates after 12.53 J/mm impact: (a) tension-dominated two-stage H-L sequence, (b) tension-dominated two-stage L-H sequence, (c) repeat H-L-H sequence.

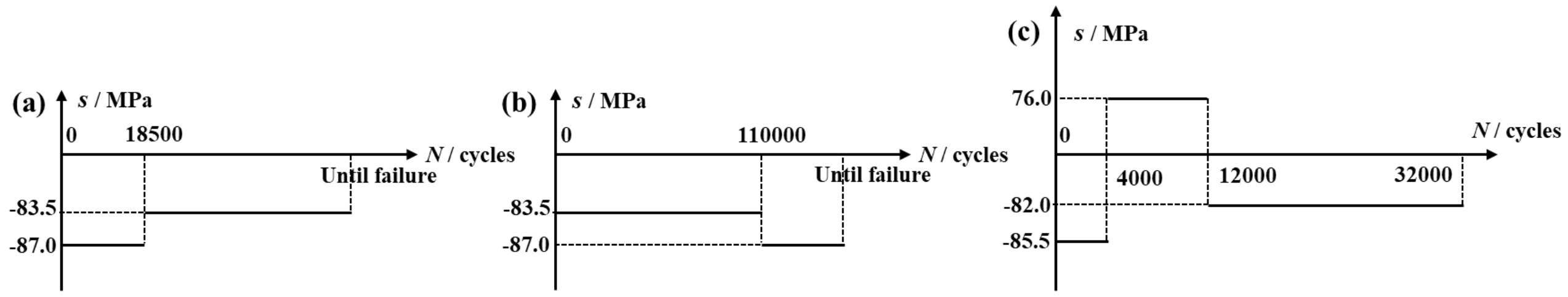

**Fig. 4** Load history of multi-step fatigue tests for $[(45/-45)/(0/90)]_{3s}$ laminates after 15.07 J/mm impact: (a) compression-dominated two-stage H-L sequence, (b) compression-dominated two-stage L-H sequence, (c) repeat H-L-H sequence.

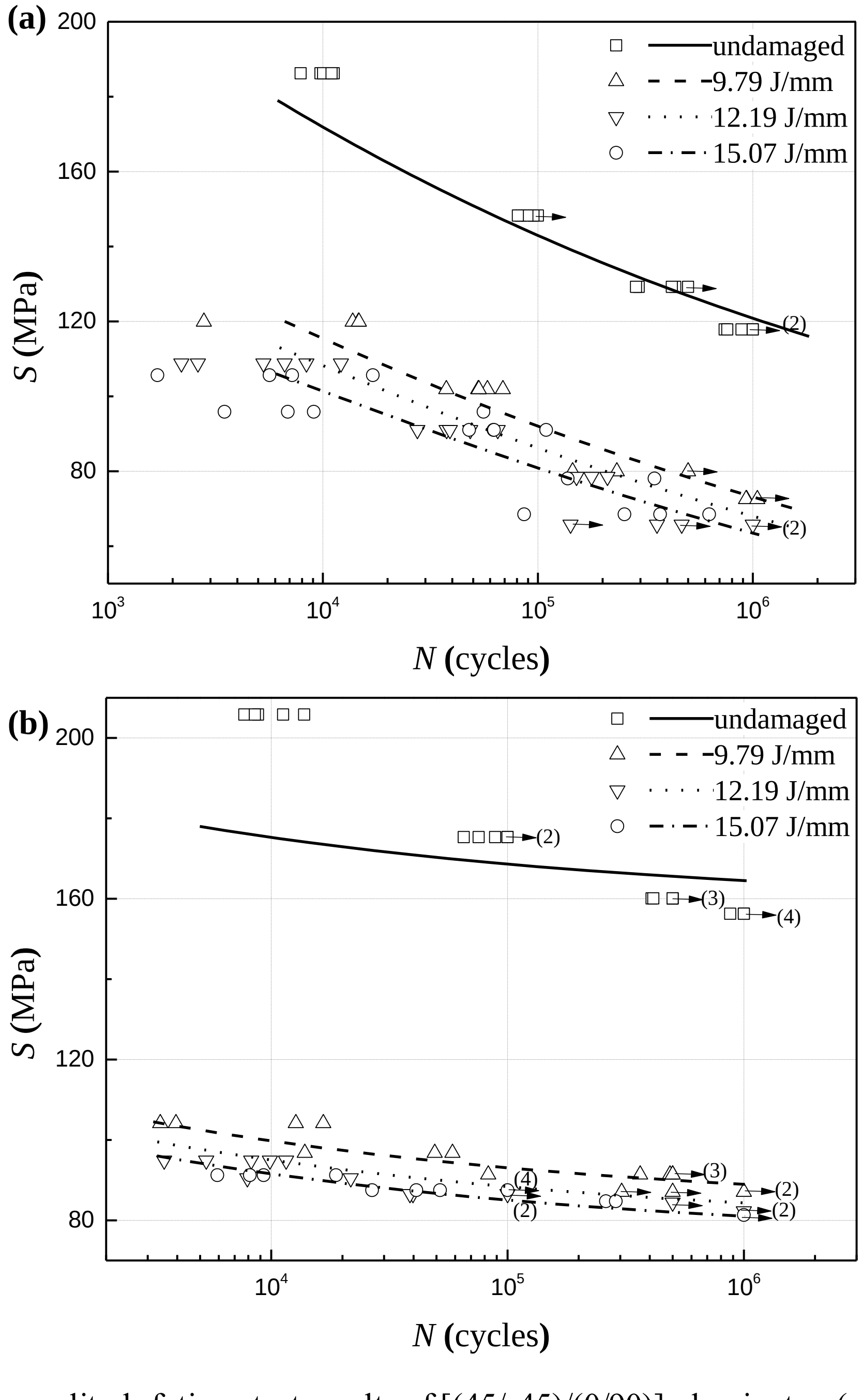

**Fig. 5** Constant-amplitude fatigue test results of $[(45/-45)/(0/90)]_{3s}$ laminates: (a) T-T fatigue load ($r = 0.05$), (b) C-C fatigue load ($r = 10$).

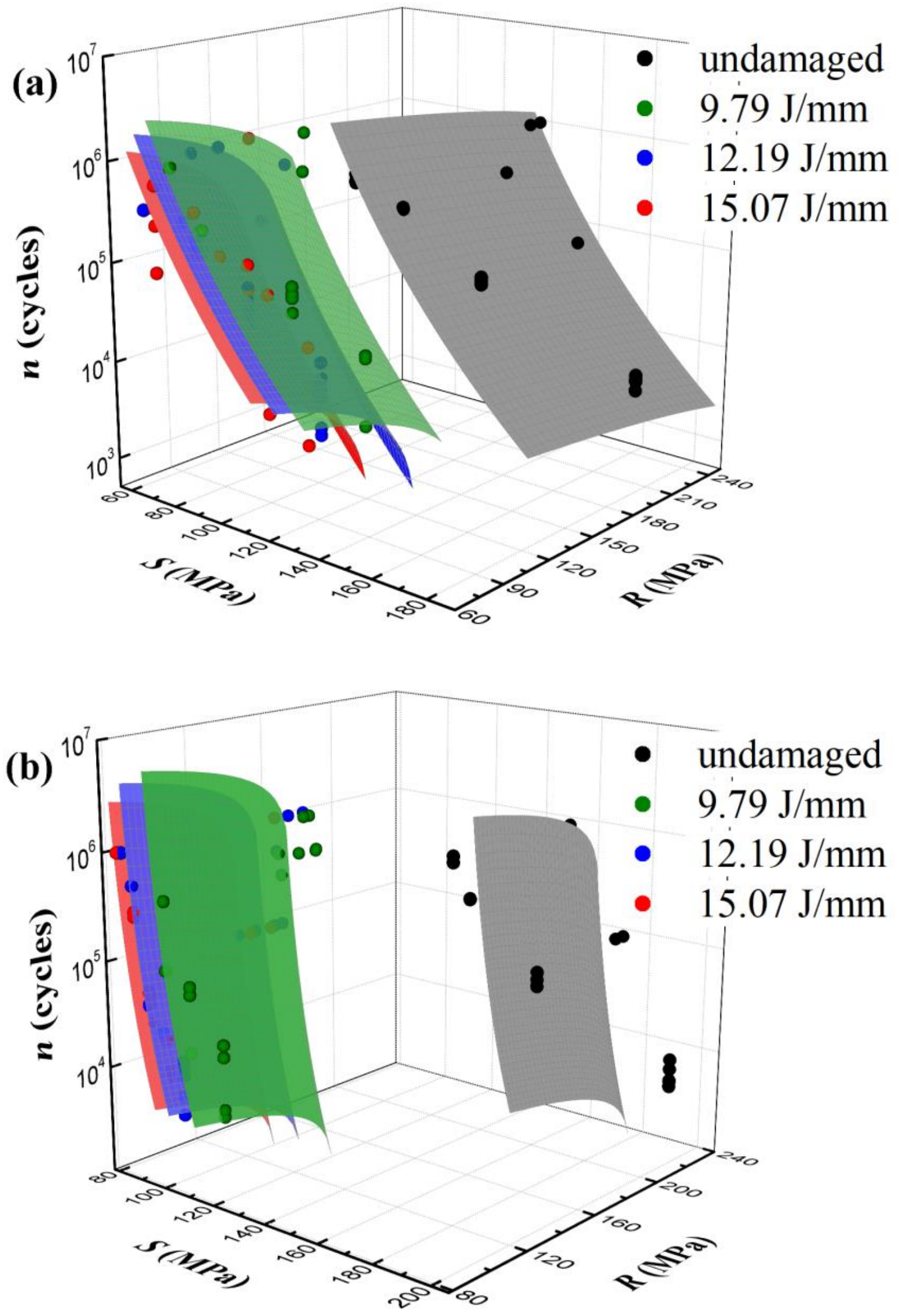

**Fig. 6** Fatigue $S$-$n$-$R$ surfaces of $[(45/\text{-}45)/(0/90)]_{3s}$ laminates: (a) T-T fatigue load ($r = 0.05$), (b) C-C fatigue load ($r = 10$).

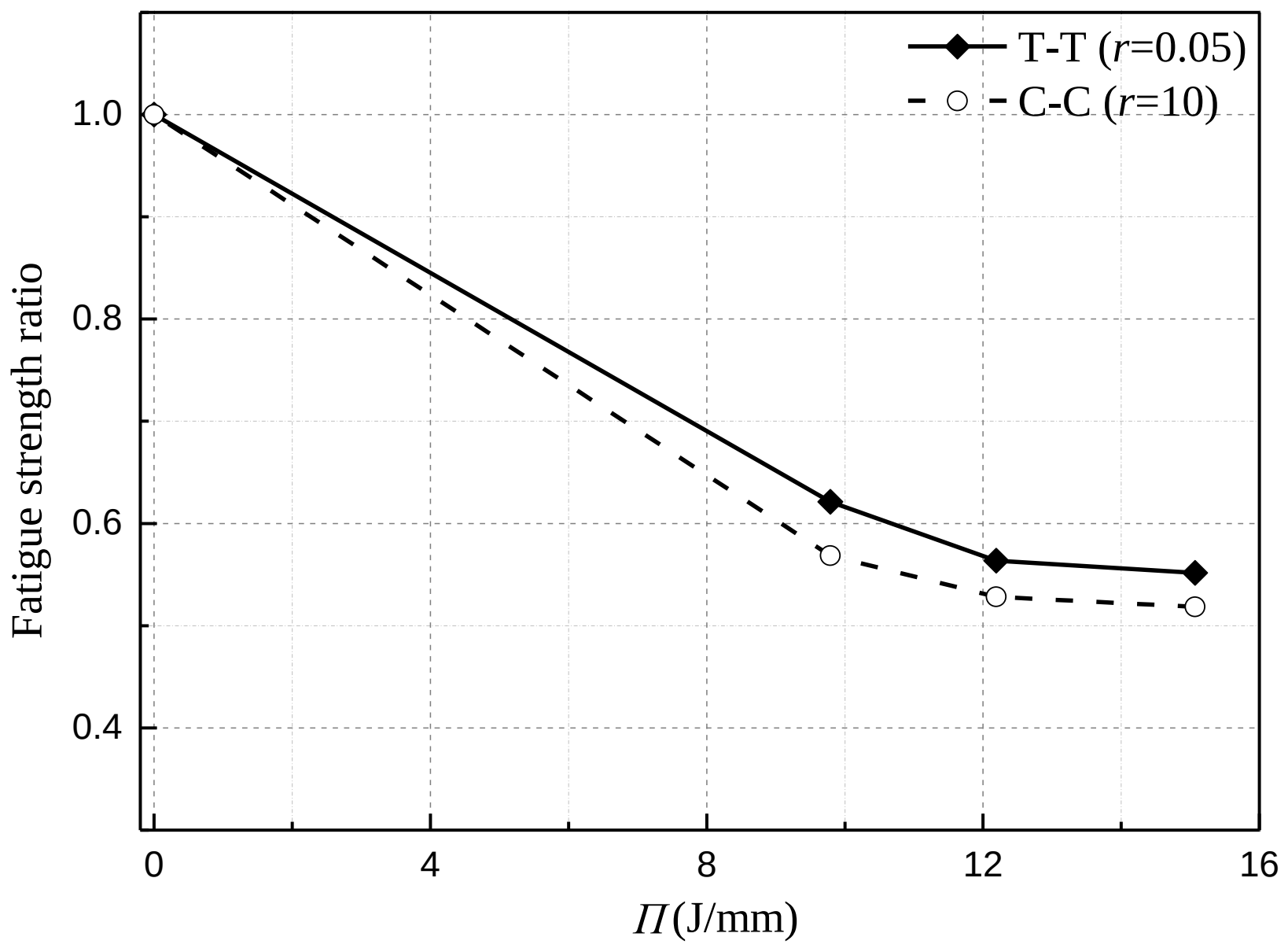

**Fig. 7** Fatigue strength ratio versus impact energy per unit thickness for $[(45/-45)/(0/90)]_{3s}$ laminates.

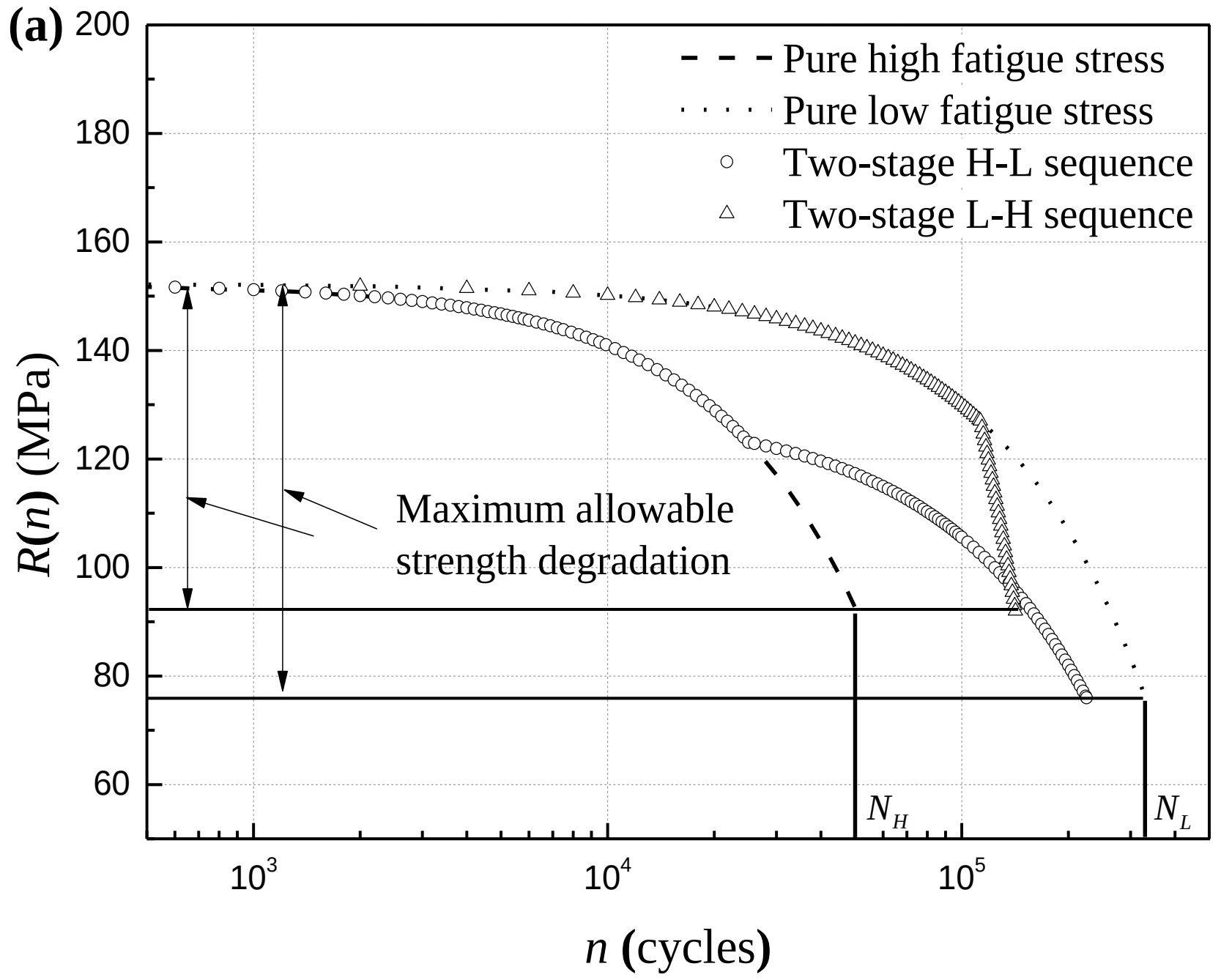

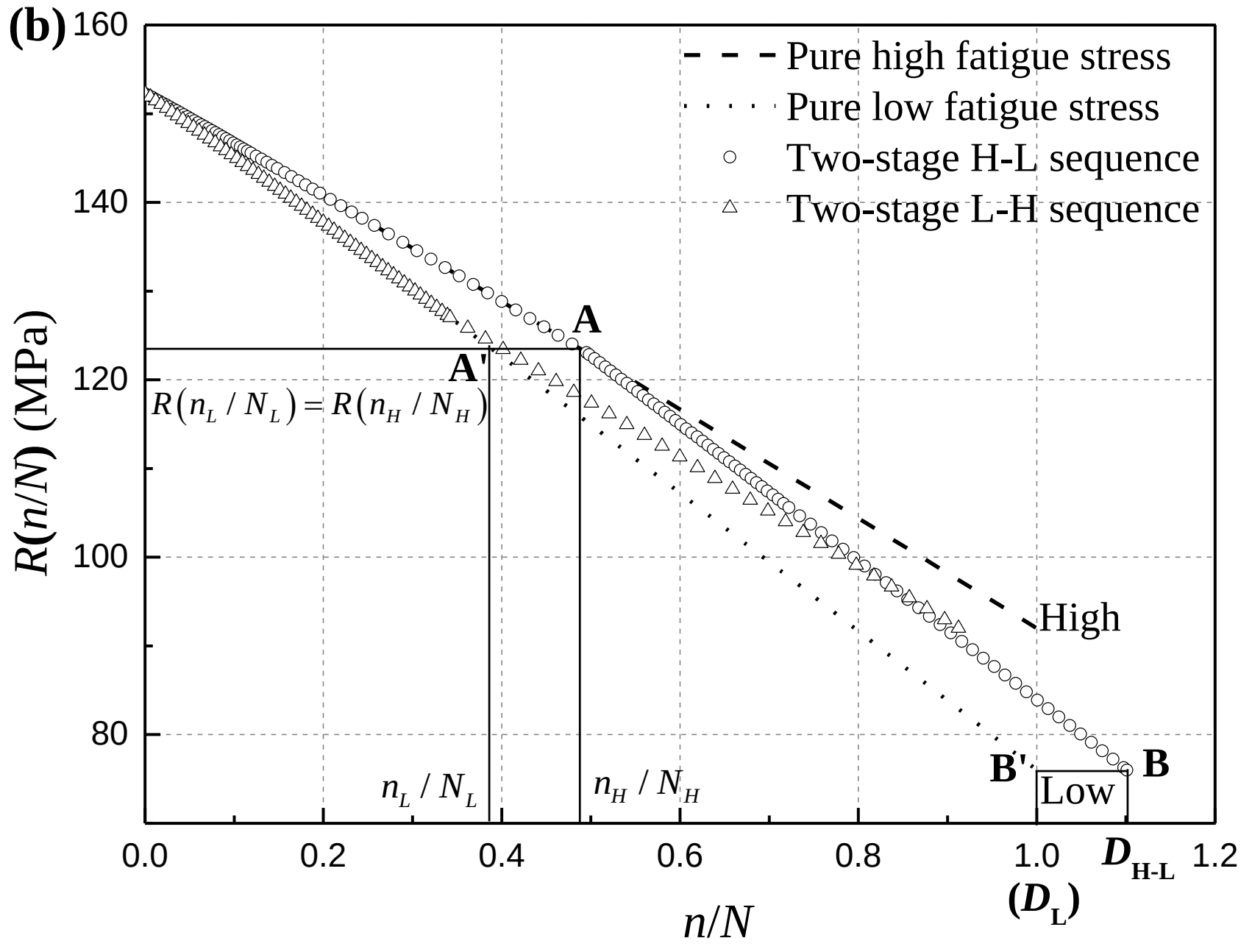

**Fig. 8** Strength degradation behaviour of $[(45/-45)/(0/90)]_{2s}$ laminates subjected to 12.53 J/mm impact under tension-dominated two-stage H-L and L-H sequences: (a) *R(n)-N*, (b) *R(n/N)-n/N*.

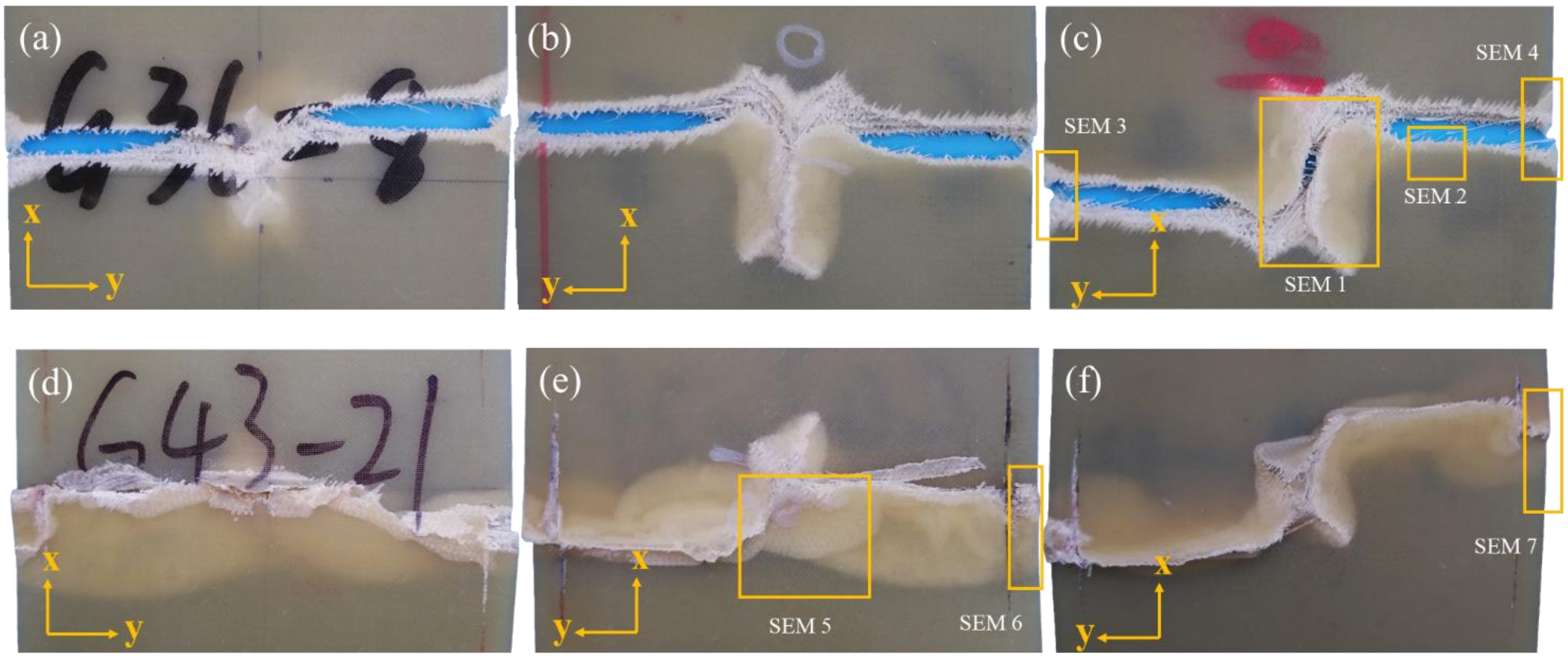

**Fig. 9** Typical macroscopic failures of impacted $[(45/-45)/(0/90)]_{3s}$ specimens under constant amplitude fatigue load (stress ratios are 0.05 and 10 for T-T and C-C fatigue loads): (a) front surface failure under T-T fatigue load, (b) back surface failure of pattern I under T-T fatigue load, (c) back surface failure of pattern II under T-T fatigue load, (d) front surface failure under C-C fatigue load, (e) back surface failure of pattern I under C-C fatigue load, (f) back surface failure of pattern II under C-C fatigue load.

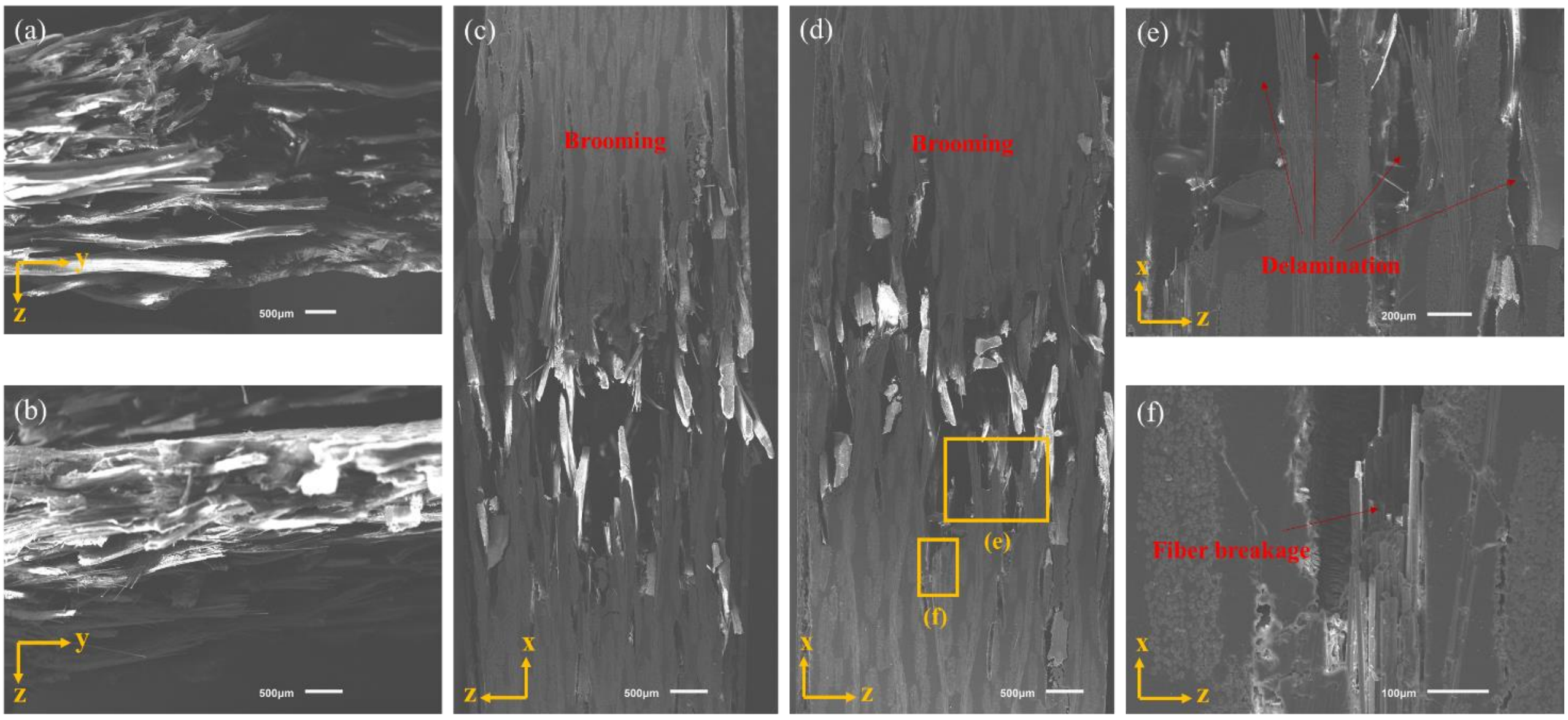

**Fig. 10** Failure modes of $[(45/\text{-}45)/(0/90)]_{3s}$ specimens under T-T fatigue load ( $r = 0.05$ ): (a) SEM 1, (b) SEM 2, (c) SEM 3, (d) SEM 4, (e),(f) magnification of (d).

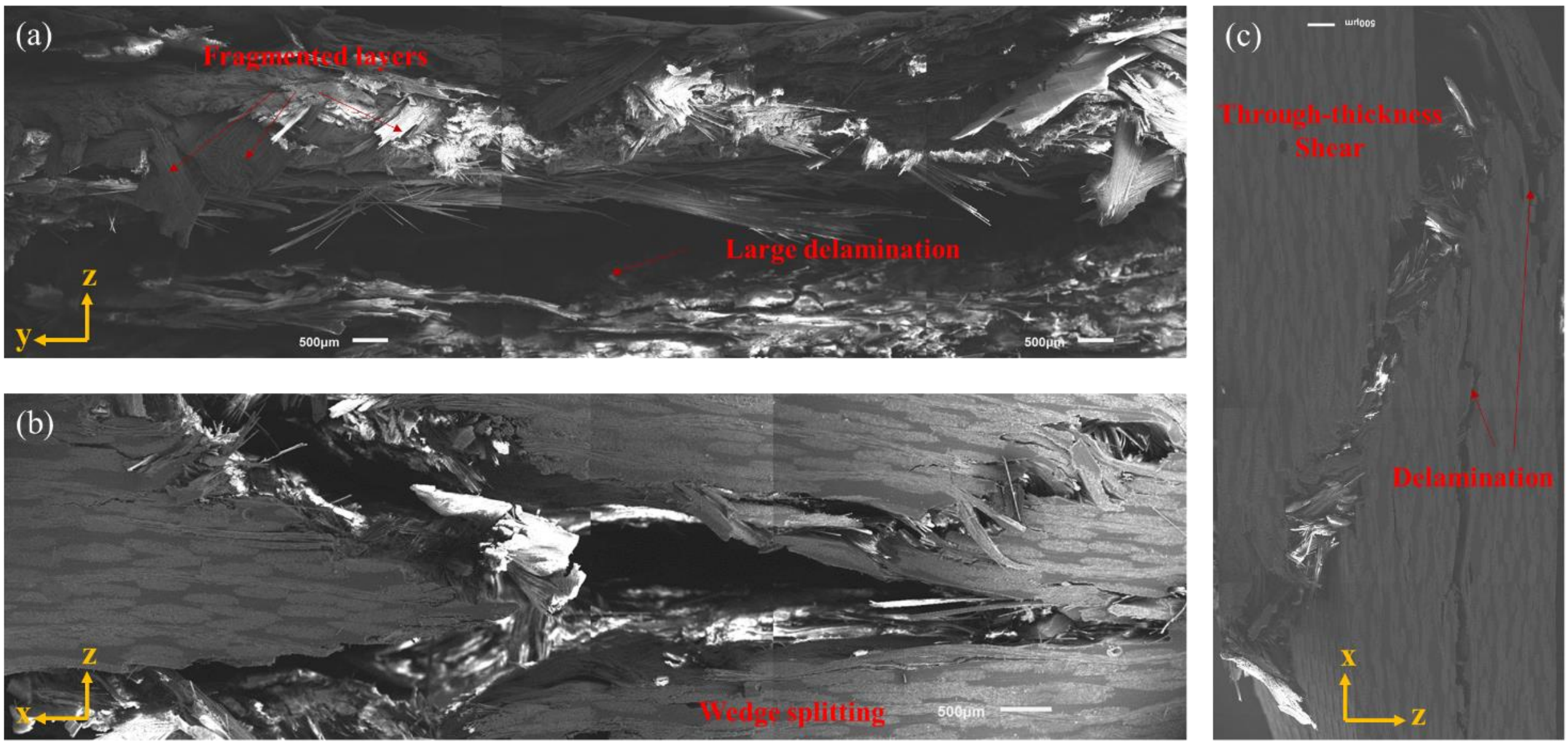

**Fig. 11** Failure modes of $[(45/\text{-}45)/(0/90)]_{3s}$ specimens under C-C fatigue load ( $r = 10$ ): (a) SEM 5, (b) SEM 6, (c) SEM 7.

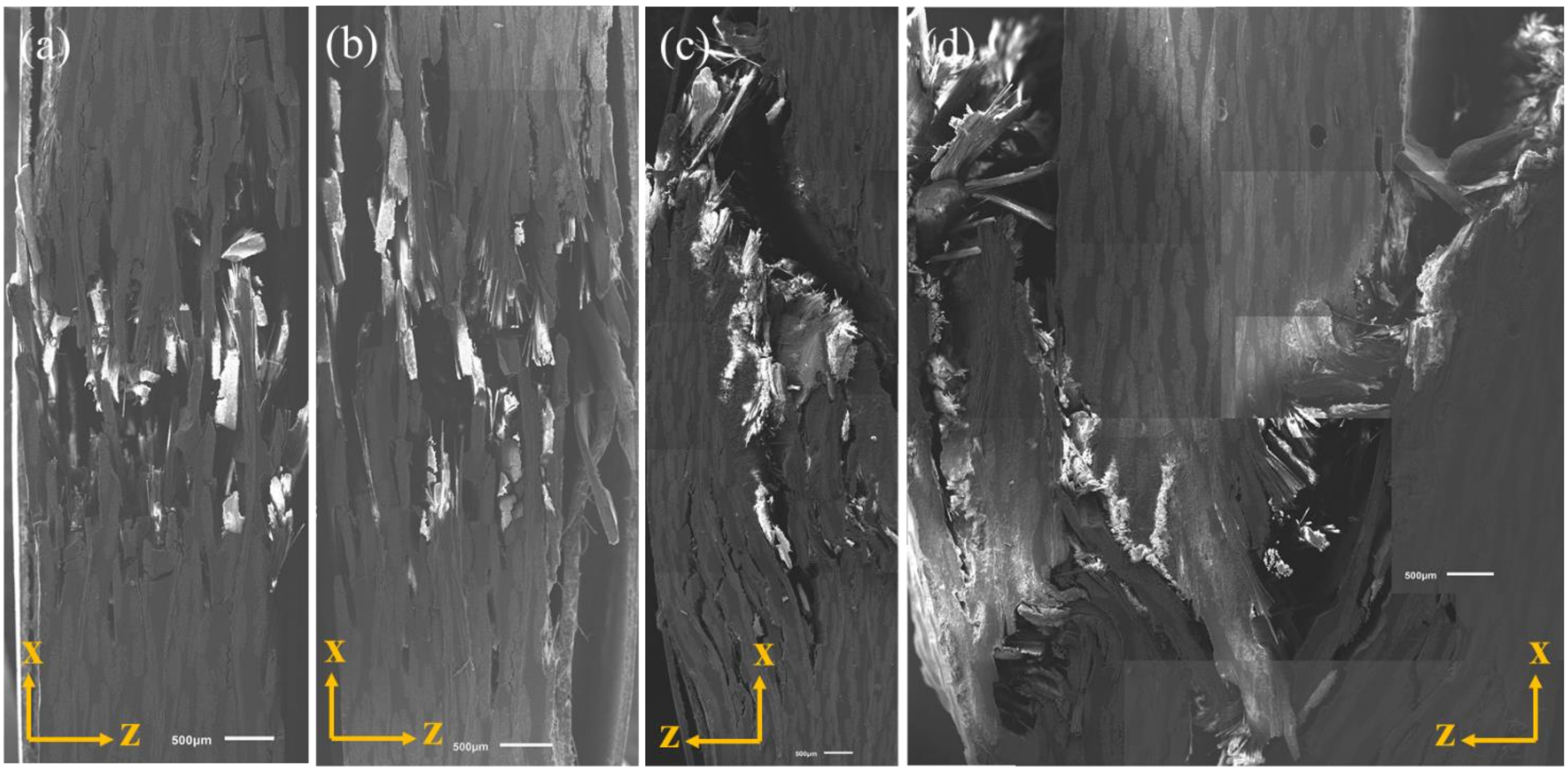

**Fig. 12** Failure modes of specimens under two-stage fatigue load: (a) tension-dominated H-L sequence for $[(45/\text{-}45)/(0/90)]_{2s}$ specimen, (b) tension-dominated L-H sequence for $[(45/\text{-}45)/(0/90)]_{2s}$ specimen, (c) compression-dominated H-L sequence for $[(45/\text{-}45)/(0/90)]_{3s}$ specimen, (d) compression-dominated L-H sequence for $[(45/\text{-}45)/(0/90)]_{3s}$ specimen.

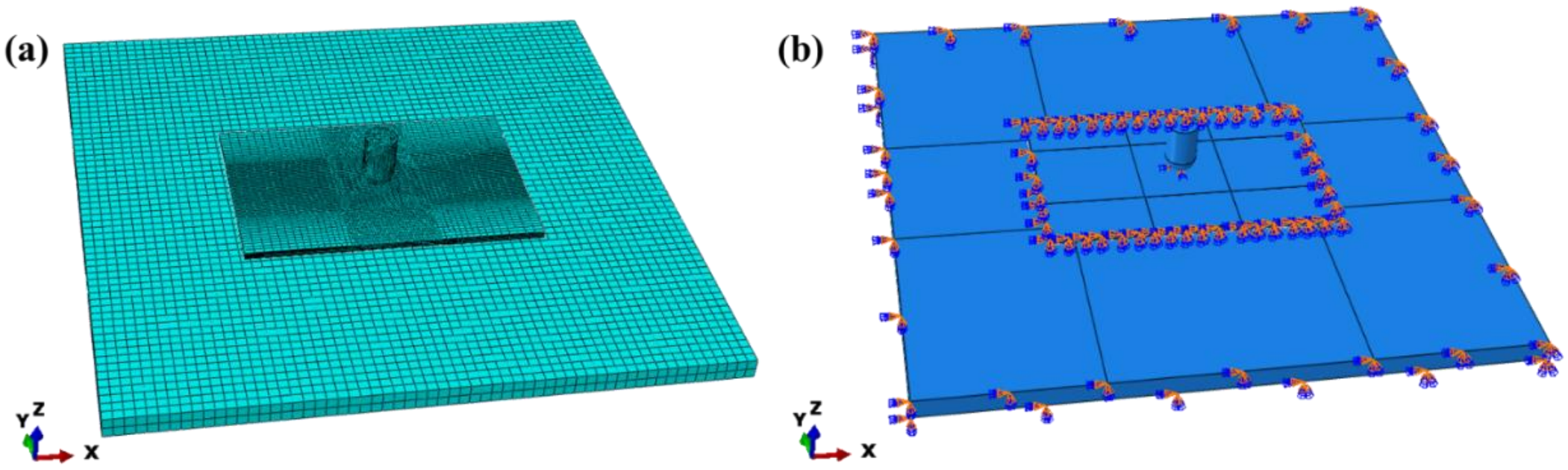

**Fig. 13** LVI finite element model: (a) FE mesh, (b) boundary conditions

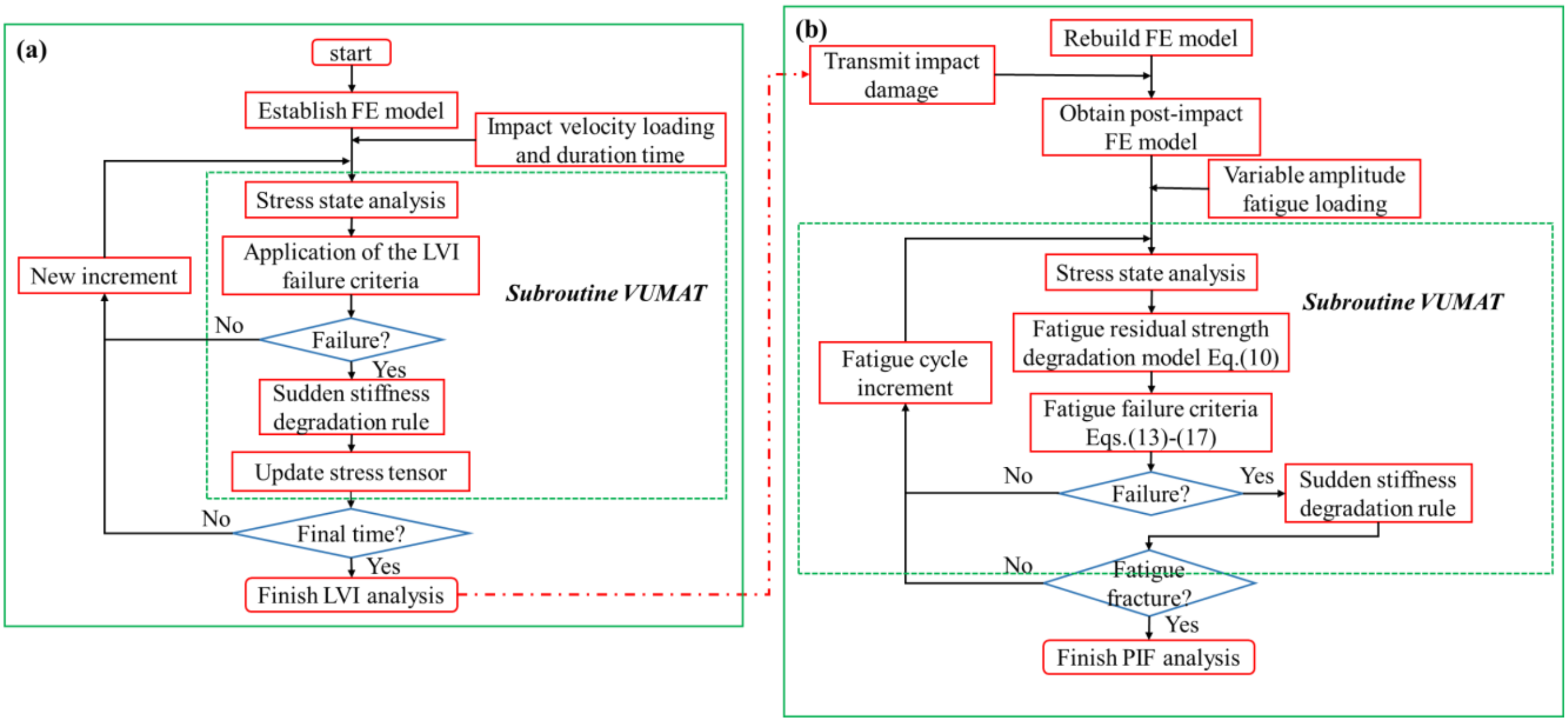

**Fig. 14** Schematic flowchart of progressive damage analysis: (a) LVI analysis, (b) PIF analysis.

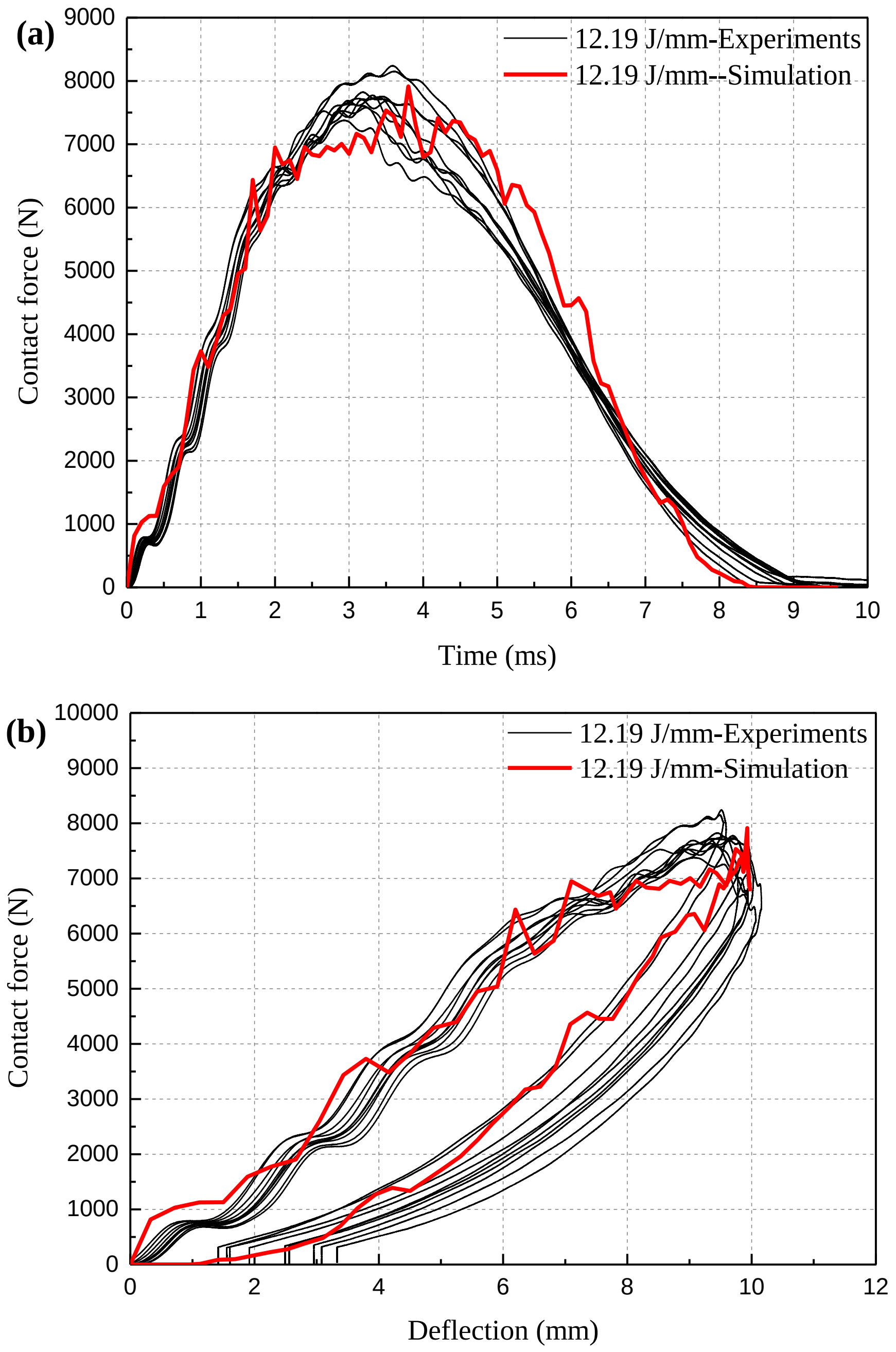

**Fig. 15** Contact force versus time and deflection curves of $[(45/-45)/(0/90)]_{3s}$ laminates at $\Pi = 12.19$ J: (a) Contact force versus time curves, (b) Contact force versus deflection curves.

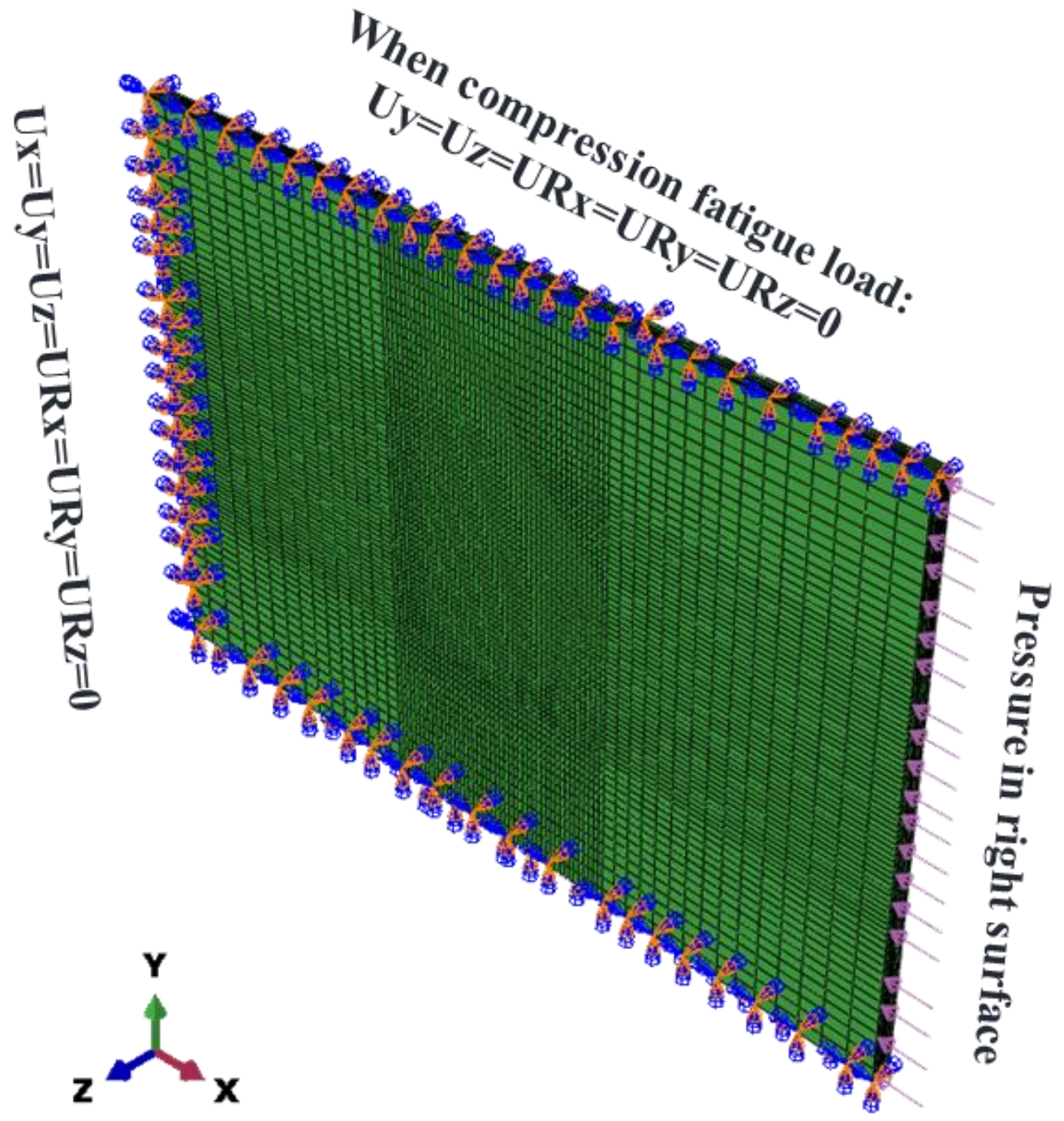

**Fig. 16** PIF finite element model

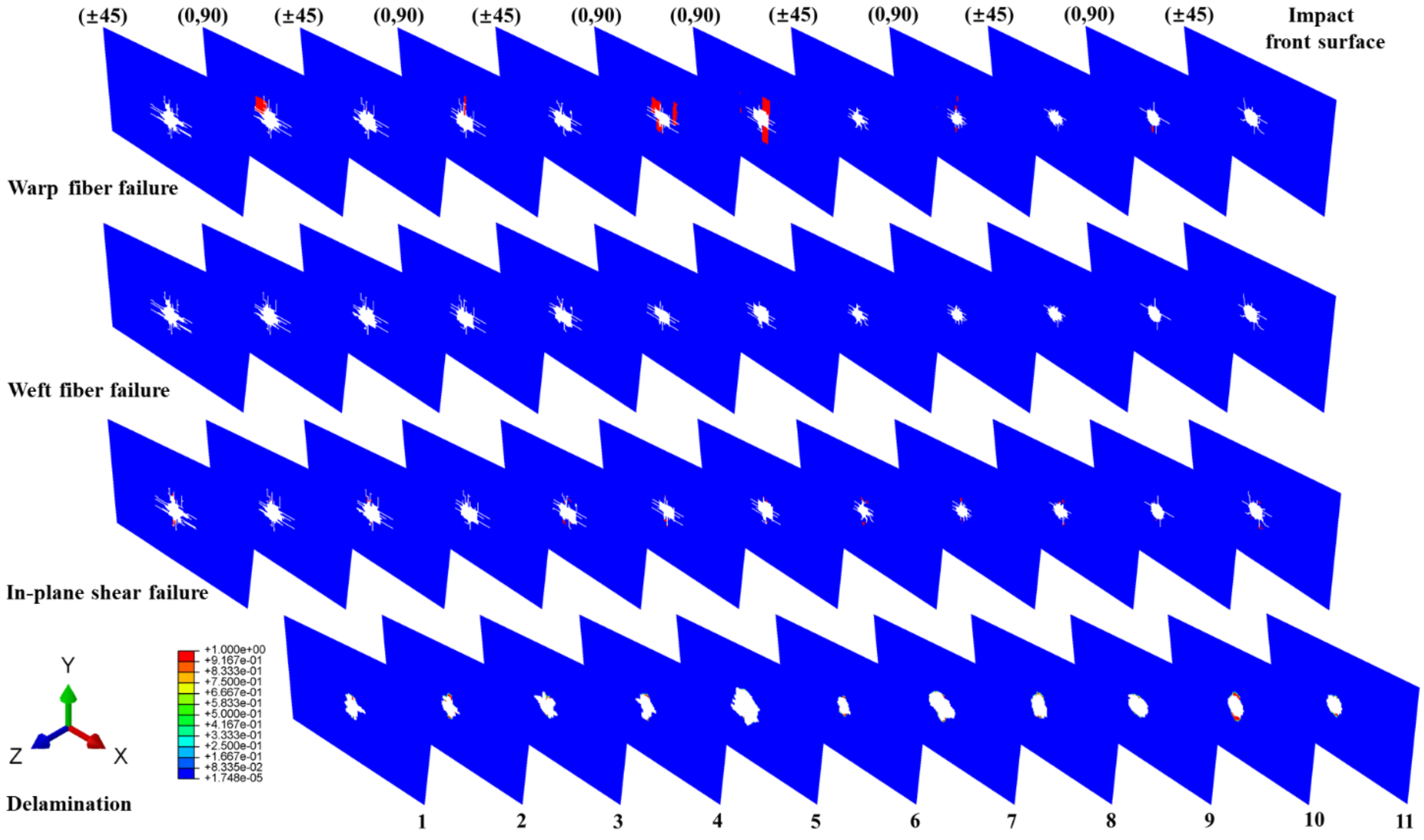

(a) 18500 fatigue cycles

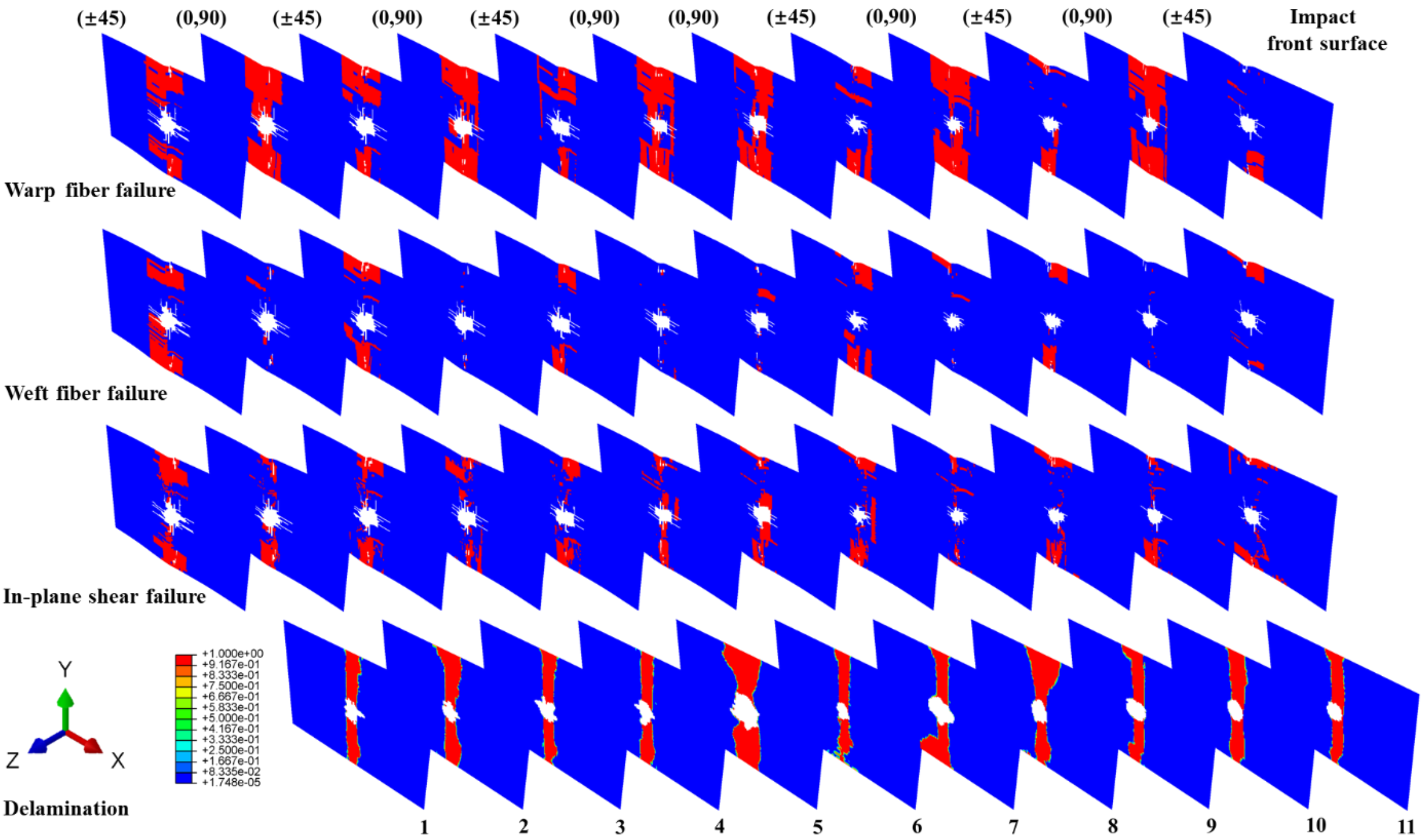

(b) 177232 fatigue cycles

**Fig. 17** Damage propagation in 15.07 J/mm post-impact $[(45/-45)/(0/90)]_{3s}$ laminate under compression-dominated two-stage H-L sequence.

**Table 1** Mechanical properties of 3238A/EW250F[21]

| $E_{1t}$ (GPa) | $E_{1c}$ (GPa) | $E_{2t}$ (GPa) | $E_{2c}$ (GPa) | $G_{12}$ (GPa) | $G_{13}$ (GPa) | $G_{23}$ (GPa) | $\nu_{12}$ |
|---|---|---|---|---|---|---|---|
| 17.8 | 22.7 | 17.7 | 24.4 | 1.7 | 1.1 | 1.1 | 0.11 |
| $X_{1t}$ (MPa) | $X_{1c}$ (MPa) | $X_{2t}$ (MPa) | $X_{2c}$ (MPa) | $X_{12}$ (MPa) | $X_{23}$ (MPa) | $V_f$ (%) | $\rho$ (g/cm$^3$) |
| 361.9 | 262.5 | 364.5 | 268.9 | 136.9 | 64.5 | 46.0 | 1.6 |

**Table 2** Multi-step fatigue test and predicted results

| $\Pi$ (J/mm) | Multi-step fatigue loading | Experimental results (cycles) | Mean Value (cycles) | $D$ | Palmgren–Miner rule (cycles) | Relative Deviation | Fatigue residual strength model (cycles) | Relative Deviation | Progressive damage Analysis (cycles) | Relative Deviation |
|---|---|---|---|---|---|---|---|---|---|---|
| 12.53 | Two-stage H-L | 255417, 175424, 213632 | 214824 | 1.07 | 191744 | 11% | 225119 | 5% | 203004 | 6% |
| | Two-stage L-H | 124031, 118411, 130135 | 124192 | 0.56 | 146220 | 18% | 141786 | 14% | 137974 | 11% |
| | Repeat H-L-H | 128187, 64030, 66858, 129993 | 97267 | 0.52 | 192257 | 98% | 128001 | 32% | 108000 | 11% |
| 15.07 | Two-stage H-L | 185641, 167868, 201904 | 185138 | 1.07 | 168286 | 9% | 171029 | 8% | 177232 | 4% |
| | Two-stage L-H | 120751, 123706, 127581 | 124013 | 0.74 | 136479 | 10% | 135803 | 10% | 133375 | 8% |
| | Repeat H-L-H | 133972, 84987, 146788 | 121916 | 0.48 | 259721 | 113% | 209120 | 72% | 143204 | 17% |

**Table 3** Material properties of cohesive elements for 3238A/EW250F[34]

| $K_{13}$ (GPa/mm) | $K_{23}$ (GPa/mm) | $K_{33}$ (GPa/mm) | $\tau_{13}$ (MPa) | $\tau_{23}$ (MPa) | $\tau_{33}$ (MPa) | $G_{\mathrm{IC}}$ (mJ/mm$^2$) |
|---|---|---|---|---|---|---|
| 430 | 430 | 1100 | 26 | 26 | 15 | 0.188 |
| $G_{\mathrm{IIC}}$ (mJ/mm$^2$) | $G_{\mathrm{IIIC}}$ (mJ/mm$^2$) | $\eta$ | Delamination initiation: $\max\left\{\frac{<\sigma_{33}>}{\tau_{33}}, \frac{\sigma_{13}}{\tau_{13}}, \frac{\sigma_{23}}{\tau_{23}}\right\}=1,\ <\sigma_{33}>=\begin{cases}\sigma_{33}, & \sigma_{33}>0\\ 0, & \sigma_{33}<0\end{cases}$ | | | |
| 3.632 | 3.632 | 1.822 | Delamination evolution: $G_{\mathrm{TC}}=G_{\mathrm{IC}}+\left(G_{\mathrm{IIC}}-G_{\mathrm{IC}}\right)\left(\frac{G_{\mathrm{II}}+G_{\mathrm{III}}}{G_{\mathrm{I}}+G_{\mathrm{II}}+G_{\mathrm{III}}}\right)^{\eta}$ | | | |

**Table 4** Sudden stiffness degradation rules[25,30]

| Failure modes | $E_{11}$ | $E_{22}$ | $E_{33}$ | $G_{12}$ | $G_{13}$ | $G_{23}$ | $\nu_{12}$ | $\nu_{13}$ | $\nu_{23}$ |
|---|---|---|---|---|---|---|---|---|---|
| Warp fibre tension failure | 0.05 | 1 | 0.05 | 0.05 | 0.05 | 1 | 0.05 | 0.05 | 1 |
| Warp fibre compression failure | 0.05 | 1 | 0.05 | 0.05 | 0.05 | 1 | 0.05 | 0.05 | 1 |
| Weft fibre tension failure | 1 | 0.05 | 0.05 | 0.05 | 1 | 0.05 | 0.05 | 1 | 0.05 |
| Weft fibre compression failure | 1 | 0.05 | 0.05 | 0.05 | 1 | 0.05 | 0.05 | 1 | 0.05 |
| In-plane shear failure | 1 | 1 | 1 | 0.05 | 1 | 1 | 0.05 | 1 | 1 |

**Table 5** Parameters of fatigue-driven residual strength model for 3238A/EW250F composites

| $r_0$ | $C_{1\mathrm{t}}$ | $p_{1\mathrm{t}}$ | $q_{1\mathrm{t}}$ | $S_{0,1\mathrm{t}}$ | $r_0$ | $C_{1\mathrm{c}}$ | $p_{1\mathrm{c}}$ | $q_{1\mathrm{c}}$ | $S_{0,1\mathrm{c}}$ |
|---|---|---|---|---|---|---|---|---|---|
| 0.05 | $7.74\times10^{9}$ | -3.51 | 0.79 | 73.21 | 10 | $2.45\times10^{15}$ | -5.60 | 0.13 | 107.84 |
| $r_0$ | $C_{2\mathrm{t}}$ | $p_{2\mathrm{t}}$ | $q_{2\mathrm{t}}$ | $S_{0,2\mathrm{t}}$ | $r_0$ | $C_{2\mathrm{c}}$ | $p_{2\mathrm{c}}$ | $q_{2\mathrm{c}}$ | $S_{0,2\mathrm{c}}$ |
| 0.05 | $7.74\times10^{9}$ | -3.51 | 0.79 | 73.21 | 10 | $2.45\times10^{15}$ | -5.60 | 0.13 | 107.84 |
| $r_0$ | $C_{12}$ | $p_{12}$ | $q_{12}$ | $S_{0,12}$ | $r_0$ | $C_{12}$ | $p_{12}$ | $q_{12}$ | $S_{0,12}$ |
| 0.05 | $8.54\times10^{5}$ | -2.16 | 0.07 | 28.96 | 10 | $8.54\times10^{5}$ | -2.16 | 0.07 | 28.96 |